\documentclass[11pt,a4paper]{article}
\pdfoutput=1
\usepackage{jcappub}
\usepackage{bm}
\usepackage{amsmath}
\usepackage{graphicx}
\def\dd{\mathrm{d}}

\def\Mpl{M_{\rm Pl}}

\newcommand{\df}{\text{d}}

\def\0{{(0)}}
\def\sig0{\dot{\sigma}_0}
\def\dvar{\delta \varphi}

\def\ph0{\dot{\phi}_0}

\title{
Generation of Primordial Black Holes and Gravitational Waves from Dilaton-Gauge Field Dynamics
}
\author[a,b]{Masahiro Kawasaki,}
\author[a,b]{Hiromasa Nakatsuka,}
\author[a]{Ippei Obata}
\affiliation[a]{ICRR, University of Tokyo, Kashiwa 277-8582, Japan}
\affiliation[b]{Kavli IPMU (WPI), University of Tokyo, Kashiwa 277-8583, Japan}
\emailAdd{obata@icrr.u-tokyo.ac.jp, hiromasa@icrr.u-tokyo.ac.jp, kawasaki@icrr.u-tokyo.ac.jp}
\abstract{
We study the observational signatures from particle production of a $U(1)$ gauge field kinetically coupled to an inflaton.
Regarding the form of gauge kinetic function, we consider the possibility that it becomes stabilized at a certain time, which makes the growing power of the gauge field evolve non-monotonically with a sharp transition.
Remarkably, the copious production of the gauge field occurs on super-horizon scales at the late stage of inflation and perturbations are enhanced on the intermediate scales during inflation.
We find that it can predict a bumpy shape of the curvature power spectrum which leads to the generation of primordial black holes as a dark matter after inflation.
We also estimate two types of tensor modes sourced by the gauge field: the primordial gravitational waves generated during inflation and the induced gravitational waves provided by the enhanced curvature perturbation after inflation.
We show that both of them are potentially testable with the future space-based gravitational wave interferometers.}

\arxivnumber{1912.XXXXX}

\usepackage{amsfonts}

\begin{document}

\begin{flushright}

\end{flushright}

\maketitle

%
%
%
\section{Introduction}

In recent decades, the precision of cosmic microwave background (CMB) and large scale structure (LSS) observations have allowed us to probe the imprint of detailed initial conditions from cosmic inflation.
One of the most important and robust predictions of inflationary cosmology is that it quantum-mechanically generates the fluctuations of spacetime called primordial gravitational waves, whose signal is quantified by the tensor-to-scalar ratio $r$ imprinted on the B-mode polarization in CMB.
The current joint collaboration of Planck and BICEP2/Keck array have constrained an upper limit of tensor-to-scalar ratio $r \lesssim 0.06$ \cite{Ade:2015lrj}.
Moreover, in 2020s, the appearance of LiteBIRD satellite \cite{Matsumura:2013aja} and CMB-S4 project \cite{Abazajian:2016yjj} will increase the sensitivity up to the order of $r = \mathcal{O}(10^{-3})$.
Throughout this measurement, we can probe the energy scale of inflation which is around the scale of grand unification theory $\sim10^{15}\text{GeV}(r/10^{-3})^{1/4}$.
Hence, if inflation is detectable in the foreseeable future, we can come to explore high energy physics utilizing the inflationary universe.

Considering the standard single-field inflationary scenarios, primordial perturbations are provided in a vacuum state and stretched out by the quasi-de-Sitter expansion of spacetime, consequently predicting (i)slightly red-tilted, (ii)isotropic, and (iii)almost Gaussian curvature power spectrum.
In addition to these, the vacuum tensor spectra are (iv)parity-symmetric.
It should be noted that, however, these statistical features (i)-(iv) are not necessarily true if the matter sector significantly contributes to the generation of perturbations in the early universe.
In the reduced 4-dimensional effective action of supergravity or string theory, for instance, the complex scalar field generically couples to the gauge field.
We conventionally call the real part of the scalar sector as the dilaton field and that of the imaginary part as the axion field.
It is well known that, once these couplings are introduced during inflation, the background motion of the scalar field can amplify the gauge quanta via the coupling function, which can source other coupled scalar or tensor perturbations on the scales where the particle production becomes relevant.
As an example, in the presence of axion-gauge coupling during inflation, a transient tachyonic instability takes place in one of the helicity modes of the gauge field around horizon crossing, whose amplification is simply proportional to the speed of rolling axion field \cite{Garretson:1992vt}.
This observable signatures in CMB have been investigated since the resultant spectrum is highly non-gaussian \cite{Barnaby:2010vf, Barnaby:2011vw, Ferreira:2014zia, Namba:2015gja, Shiraishi:2016yun, Agrawal:2018mrg, Dimastrogiovanni:2018xnn, Fujita:2018vmv}.
Also, intriguingly, the tensor mode sourced by the helical gauge field is parity-violated due to the helicity conservation \cite{Sorbo:2011rz, Dimastrogiovanni:2012ew, Adshead:2013qp}.
Moreover, the resultant spectral shapes can be either blue or red, or can even have bumps being determined by the form of axion's potential.
As a consequence, it also predicts the observables at intermediate scales such as the generation of the primordial black holes (PBHs) and the scale-dependent gravitational waves, which could be detected with the upcoming measurements in CMB, pulsar-timing arrays or gravitational wave detectors \cite{Cook:2011hg, Anber:2012du, Barnaby:2012xt, Linde:2012bt, Mukohyama:2014gba, Obata:2014loa, Obata:2016tmo, Domcke:2016bkh, Maleknejad:2016qjz, Guzzetti:2016mkm, Obata:2016xcr, Dimastrogiovanni:2016fuu, Adshead:2016omu, Garcia-Bellido:2016dkw, Obata:2016oym, Fujita:2017jwq, Garcia-Bellido:2017aan, Thorne:2017jft, Ozsoy:2017blg, Fujita:2018ndp}.

In this paper, we focus on another gauge sector coupled to the dilaton field and develop its cosmological phenomena caused by the gauge field production.
It is known that the background motion of the dilaton field breaks the conformal invariance of the electromagnetic action via the time variation of kinetic coupling function,
which can occur an instability in both helicity modes of the gauge field on super-horizon scales.
Since the vector field with longer coherent length is more enhanced while the instability persists, this mechanism has been motivated to explain the observables over kpc-Gpc scales such as an intergalactic magnetic field \cite{Ratra:1991bn, Martin:2007ue, Demozzi:2009fu, Kanno:2009ei, Durrer:2010mq, Fujita:2012rb, Fujita:2013pgp, Fujita:2014sna, Obata:2014qba, Fujita:2016qab, Caprini:2017vnn, Fujita:2019pmi} or the possible of statistical anisotropies in CMB from the model of anisotropic inflation \cite{Watanabe:2009ct, Watanabe:2010fh, Kanno:2010nr, Watanabe:2010bu, Do:2011zza, Soda:2012zm, Bartolo:2012sd, Ohashi:2013qba, Ohashi:2013pca, Naruko:2014bxa, Choi:2015wva, Abolhasani:2015cve, Ito:2017bnn, Fujita:2017lfu, Fujita:2018zbr, Hiramatsu:2018vfw}.
Contrary to these attempts for the phenomenological interest on large scales, however, these observational signatures on the smaller scales have been less clarified.
One of the main reasons is the fact that a specific class of coupling function has been assumed,
which leads to a constant growth power of the gauge field.
As a result, the perturbations on the largest scale are naturally amplified during inflation.
Actually, it will not be necessary to impose such a relation.
Emphasizing this point, in this work we study the phenomenology of particle production from the dilaton-induced gauge field occurring at the late stage of inflation.
As a previous work, Ito \& Soda have studied the inflationary model with an exponential type gauge kinetic function where the gauge field starts to grow in the last stage of inflation and predicted the sourced primordial gravitational waves in the MHz frequency band \cite{Ito:2016aai}.
In their scenario, however, the form of coupling function predicts a monotonic increasing growth power of the gauge field so that the period of particle production is severely constrained to be close to the end of inflation.
We extend their minimum model and newly introduce a constant term to the gauge kinetic function.
This assumption leads to the non-monotonic time evolution of the growing power of gauge field, and consequently, it triggers a short-term particle production of the gauge field and enhances other coupled fluctuations on the intermediate scales during inflation.
As a first step, we develop the possibility of generating PBHs as a dark matter sourced by the gauge field.
It is well known that PBHs are formed if the high-density regions reenter the horizon after inflation and collapse gravitationally in the radiation dominated era \cite{Hawking:1971ei, Carr:1974nx, Carr:1975qj}.
The PBH formation requires large perturbations on a small scale which can be achieved by the extension of the inflation model \cite{Garcia-Bellido:2016dkw, Bugaev:2011wy, Inomata:2017vxo, Kohri:2012yw, Ando:2017veq, Pi:2017gih, Ando:2018nge}.
In our model, we find that the power spectrum of curvature perturbation sourced by the gauge field can explain a sizable amount of present dark matter as PBH.
In addition to the scalar mode, we also estimate two types of tensor modes associated with the gauge field production.
One is the tensor mode sourced by the gauge field during inflation and we call it  ``primordial gravitational waves".
The other is the tensor mode sourced by the enhanced curvature perturbation re-entering the horizon during the radiation-dominated era after inflation.
We call it  ``induced gravitational waves".
We show that both of the resultant power spectra are testable with the future space-based laser interferometers such as DECIGO \cite{Kawamura:2011zz}, BBO \cite{Crowder:2005nr}, or potentially LISA \cite{Audley:2017drz} missions.

This paper is organized as follows.
In section \ref{Model Action and Setup}, we set up our model and demonstrate how the gauge field is amplified on super-horizon scales in our scenario.
Next, we calculate the power spectrum of scalar mode sourced by the gauge field and derive the resultant mass spectrum of PBHs in section \ref{Perturbation Dynamics}.
In section \ref{tensor}, we estimate the power spectra of both primordial and induced gravitational waves and present their detectability.
In section \ref{discussion}, we discuss a couple of computational and observational consistencies in our model.
Finally, we summarize our work and discuss the outlook in section \ref{conclusion}. In this paper, we set the natural unit $\hbar = c = 1$.

\section{Particle production of gauge field and background dynamics}
\label{Model Action and Setup}

In this section, we present how the particle production of gauge field occurs in our background dynamics.
We set up the following Lagrangian density
\begin{equation}
\mathcal{L} = \dfrac{\Mpl^2}{2}R - \frac{1}{2}(\partial_\mu \varphi)^2 -V(\varphi)
- \frac{1}{4}I^2(\varphi)F_{\mu\nu}F^{\mu\nu},
\label{model action}
\end{equation}
where $R$ is the Ricci scalar, $\Mpl$ is the reduced Planck mass, $\varphi$ is the inflaton with its potential $V(\varphi)$ and $F_{\mu\nu} = \partial_\mu A_{\nu} - \partial_\nu A_{\mu}$ is the field strength of gauge field $A_\mu$.
The inflaton field $\varphi$ is coupled to the kinetic term of
the gauge field via $I(\varphi)$.
We decompose these fields into the backgrounds and perturbations as
\begin{align}
\varphi(t,\bm{x})=\bar{\varphi}(t)+\dvar(t,\bm{x}), \quad A_{i}(t,\bm{x}) = \delta A_{i}(t, \bm{x}) \ ,
\end{align}
where for the gauge field we take $A_0(t,\bm{x}) = \partial_i A_{i}(t,\bm x) = 0$.
Throughout this paper, we assume that the gauge field has no classical homogeneous vector field and the produced gauge field fluctuations on large scales have a negligible effect on the background dynamics.
Hence, regarding the metric form, we adapt the Friedmann-Robertson-Walker background $ds^2 = -dt^2 + a(t)^2d\bm{x}^2 = a(\tau)^2(-d\tau^2 + d\bm{x}^2)$ without the spatial anisotropy in its component.

To understand how the particle production occurs, let us quantize the electromagnetic field and decompose it into the linear polarization vectors $e^{X}_i(\hat{\bm{k}})$ and $e^{Y}_i(\hat{\bm{k}})$ (see appendix \ref{ap: p} for their definition) in Fourier space
\begin{align}
&A_i(t, \bm{x}) = \int\dfrac{d\bm{k}}{(2\pi)^3}\left( \hat{A}^X_{\bm{k}}(t)e^X_i(\hat{\bm{k}}) + i\hat{A}^Y_{\bm{k}}(t)e^Y_i(\hat{\bm{k}})\right)e^{i\bm{k}\cdot\bm{x}} \ , \\
&\hat{A}^\lambda_{\bm{k}}(t) = A^\lambda_k(t)a^\lambda_{\bm{k}} + A^{\lambda*}_k(t)a^{\lambda\dagger}_{-\bm{k}} \ , \qquad \left[ a^\lambda_{\bm{k}}, \ a^{\lambda'\dagger}_{-\bm{k}'} \right] = (2\pi)^3\delta^{\lambda\lambda'}\delta(\bm{k}+\bm{k}') \quad (\lambda = X, \ Y) \ ,
\end{align}
where $\{a^\lambda, a^{\lambda\dagger}\}$ are quantum annihilation/creation operators satisfying the ordinal commutation relations.
Then, defining the dimensionless time variable $x \equiv -k\tau$, the equation of motion for $A^\lambda_k$ is given by
\begin{equation}
\left[\partial_x^2 +  1 - \dfrac{\partial_x^2 \bar I}{\bar I} \right](\bar{I}A_k) = 0 \ , \label{eq: gauge}
\end{equation}
where $\bar I \equiv I(\bar \varphi)$ and we omitted the polarization index $\lambda$ since both polarization modes obey the same equation of motion.
We can see that the time variation of the gauge kinetic function explicitly violates the conformal invariance of gauge field. 
For our analytical convenience, we denote the growth rate of $I(\varphi)$ as
\begin{equation}
n(t) \equiv \dfrac{d\ln \bar I}{d N} = -\dfrac{\bar I_\varphi}{\bar I}\dfrac{\dot{\bar \varphi}}{H} \ ,
\end{equation}
where we defined the number of e-foldings as $dN \equiv -Hdt$.
In this work, we consider only the case as $n(t)$ is positive: $n(t) > 0$.
Since $n(t)$ varies slowly, \eqref{eq: gauge} approximately leads to
\begin{equation}
\left[\partial_x^2 +  1 - \dfrac{\nu(t)^2-1/4}{x^2} \right](\bar{I}A_k) \simeq 0 \ , \qquad \nu(t)^2 = (n(t)-1/2)^2 \ . \label{eq: gauge2}
\end{equation}
%
If $\nu(t)$ ($n(t)$) is almost constant in the whole period of inflation,  the solution of \eqref{eq: gauge2} with the Bunch-Davies initial condition is well described by the Hankel function of the first kind
\begin{equation}
 \bar{I} A_k(x) = \dfrac{e^{i\tfrac{2\nu+1}{4}\pi}}{\sqrt{2k}}\sqrt{\dfrac{\pi x}{2}}H^{(1)}_\nu(x) \label{eq: Hankel} \ .
 \end{equation}
Here, we can choose $\nu \geq 0$ depending on the value of $n$.
Then using the following asymptotic form
\begin{equation}
H^{(1)}_{\nu>0}(x) \rightarrow -\dfrac{i}{\pi}\Gamma(\nu)\left(\dfrac{2}{x}\right)^\nu \ , \quad H^{(1)}_0(x) \rightarrow \dfrac{2i}{\pi}\ln x \qquad (x\rightarrow 0) \ , 
\end{equation}
the mode functions of electromagnetic field on super-horizon scales are expressed as
%
%
%
\begin{equation}
E_k = \dfrac{\bar I \dot{A}_k}{a} \rightarrow
\begin{cases}
-ie^{i\frac{n}{2}\pi}\dfrac{4H^2\Gamma(n+\tfrac{1}{2})}{\sqrt{2\pi k^3}}\left(\dfrac{x}{2}\right)^{2-n} & (n \geq 2) \\
0 & (0 < n < 2)
\end{cases} \qquad (x \rightarrow 0) \ .
\end{equation}
\begin{equation}
B_k = \dfrac{k\bar{I}A_k}{a^2} \rightarrow
\begin{cases}
-ie^{i\frac{n}{2}\pi}\dfrac{4H^2\Gamma(n-\tfrac{1}{2})}{\sqrt{2\pi k^3}}\left(\dfrac{x}{2}\right)^{3-n} & (n \geq 3) \\
0 & (0 < n < 3)
\end{cases} \qquad (x \rightarrow 0) \ .
\end{equation}
Therefore the electric energy density $\rho_E \equiv \bar I^2\dot{A}_i^2/(2a^2)$ blows up on super-horizon scales if $n \geq 2$ is satisfied.
Hereafter, we disregard the contribution from the magnetic field since it is always subdominant compared to that from the electric field.

\begin{figure}[tbp]
\center
  \includegraphics[width=80mm]{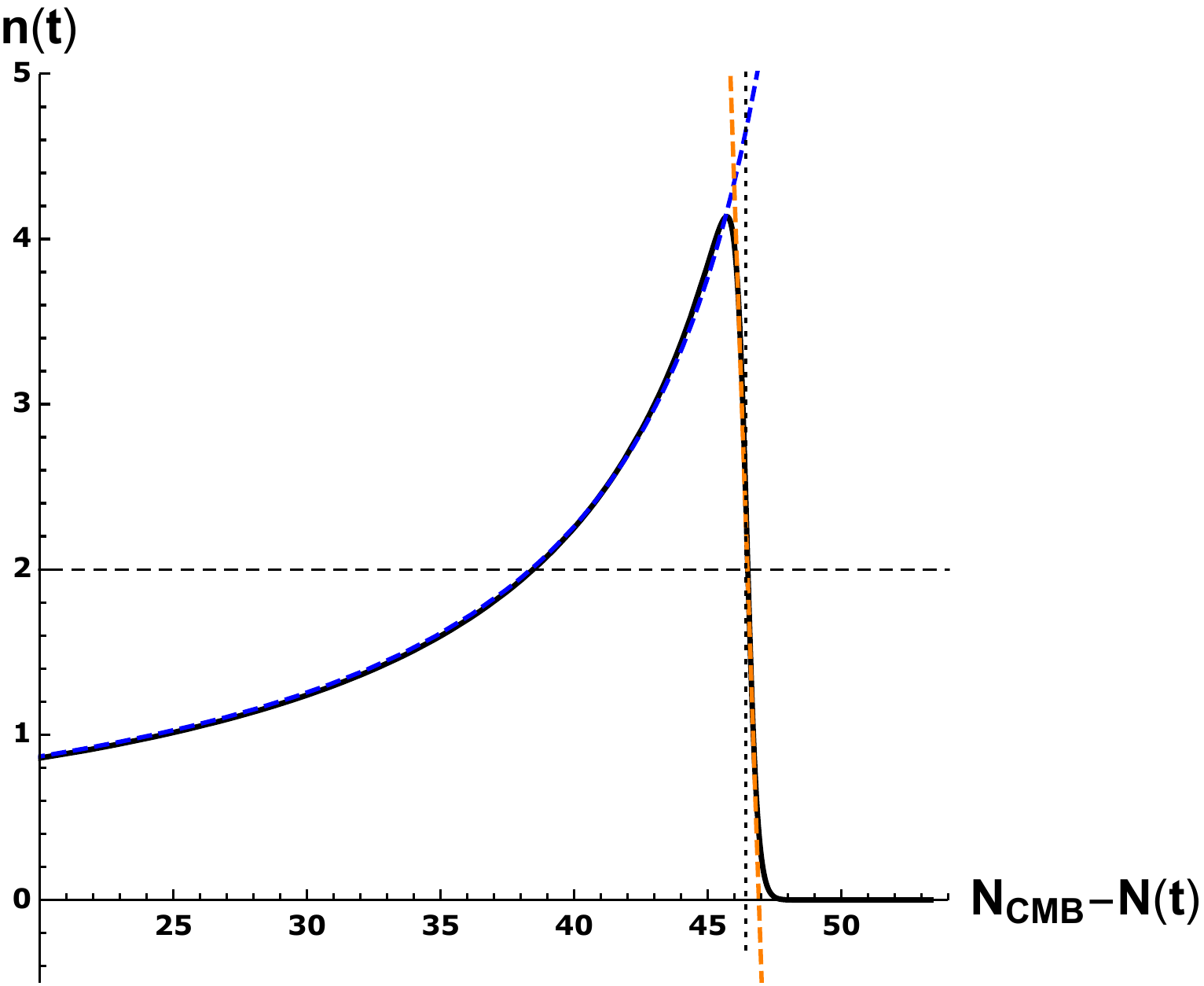}
  \caption{
  Time evolution of $n(t)$ with $\{B_2/B_1\simeq 8.8\times10^{25}, \  c_1\simeq23\}$.
  The horizontal axis is the number of e-foldings $N_{\rm CMB}-N(t) = \ln (a(t)/a_{\rm CMB})$.
  The black solid line shows the numerical evolution.
  The dashed blue and orange lines are the approximate solutions \eqref{eq: na} with $\{\alpha=1.5, \ \beta=1.2\}$.
  At first, $n(t)$ is monotonically increasing until around the transition time $\varphi(t_t) = \varphi_t$.
  After that, it quickly decreases and inflation ends a few e-folds later.
  }
 \label{fig:n}
\end{figure}

For analytical convenience, a specific class of functional form $I(\varphi)$ has been assumed such as $I(\varphi) \propto \exp(c\int d\varphi V/V_\varphi)$ which realizes the above analytical solution (namely $n(t) \simeq \text{const.}$).
We revisit such a treatment and let us consider the possibility that we have a following functional form
%
\begin{equation}
I(\varphi) = B_1\exp{\left(c_1\dfrac{\varphi}{M_{\rm Pl}}\right)} + B_2 \ , \label{eq: I}
\end{equation}
where $B_1, \ B_2, \ c_1 > 0$ are model parameters.
In addition to a conventional exponential function, we also introduce another constant term expected to arise from a string-loop modification in powers of a dilaton-dependent coupling constant \cite{Damour:1994zq}.
We assume that in \eqref{eq: I} the first term is sufficiently greater than the second term at initial stage, and accordingly $n(t)$ is monotonically increasing as $\bar{\varphi}(t)$ is decreasing in time.
However, at a certain time $n(t)$ stops growing and a transitional behavior happens when both terms get balanced.
%
%
%
In order to see it, we solve the Friedmann equation and the background equation of motion for $\bar\varphi$ given by
\begin{align}
&3\Mpl^2H^2 = V + \langle\rho_E\rangle \ , \\
&\ddot{\bar{\varphi}} + 3H\dot{\bar{\varphi}} + V_\varphi = \dfrac{2\bar{I}_\varphi}{\bar{I}}\langle\rho_E\rangle \ ,
\end{align}
where
\begin{align}
\langle\rho_E(\tau)\rangle &= \int d\ln k ~\dfrac{k^3}{2\pi^2}|E_k(\tau)|^2 \label{eq: eleenergy}
\end{align}
is the backreaction of gauge field.
Regarding the potential $V(\varphi)$, we adopt the Starobinsky-type potential for instance\footnote{In string theory, a similar potential form to Starobinsky inflation is known to be realized by a presence of D-brane defect \cite{Ellis:2014cma}. We also emphasize that the choice of specific potential form is not sensitive to our result.}
\begin{align}
V(\varphi)&= \mu^4\left( 1 - e^{-\gamma \varphi} \right)^2  \ , \qquad \gamma = \sqrt{\dfrac{2}{3}}M_{\rm {Pl}}^{-1} \ .
\label{potential V}
\end{align}
Throughout this paper we fix the values of the energy scale $\mu$ and the field range of $\bar{\varphi}$ on the CMB scale $\bar\varphi(t_{\rm{CMB}}) \equiv \bar\varphi_{\rm{CMB}}$ in order to satisfy the CMB constraint.
Without the backreaction, we obtain the following slow-roll relations
\begin{equation}
\dfrac{d\bar \varphi}{dN} = -\dfrac{\dot{\bar \varphi}}{H} =\dfrac{2M_{\rm Pl}^2~\gamma ~e^{-\gamma \bar\varphi}}{1 - e^{-\gamma \bar\varphi}} \qquad \longleftrightarrow \qquad N(t) \simeq \dfrac{e^{\gamma \bar\varphi(t)}}{2M_p^2 \gamma^2} \qquad (\gamma\bar\varphi \gg 1) \end{equation}
%
and then $n(\bar\varphi(t))$ is approximately represented by
\begin{align}
n(\bar\varphi) &= -\dfrac{\dot{\bar{\varphi}}}{HM_{\rm Pl}}\dfrac{B_1c_1\exp{\left(c_1\dfrac{\bar\varphi}{M_{\rm Pl}}\right)}}{B_1\exp{\left(c_1\dfrac{\bar\varphi}{M_{\rm Pl}}\right)} + B_2} \\
&\simeq \dfrac{1}{M_{\rm Pl} \gamma N}\dfrac{B_1c_1\left(2M_{\rm Pl}^2 \gamma^2N\right)^{\tfrac{c_1}{\gamma M_{\rm Pl}}}}{B_1\left(2M_{\rm Pl}^2 \gamma^2N\right)^{\tfrac{c_1}{\gamma M_{\rm Pl}}} + B_2} \ . \label{eq: n1}
\end{align}
We plot the time evolution of $n(t)$ in Figure \ref{fig:n} with respect to the number of e-foldings $N_{\rm CMB} - N$. $N_{\rm CMB}$ is the number of e-folding on the CMB scale and here we chose $N_{\rm CMB} \sim 50$ in this plot.
At first, $n(t)$ monotonically increases and becomes greater than 2 at around 40 number of e-foldings.
The first term of \eqref{eq: I} is relevant until the time $t = t_m$ when $n$ gets a maximum value $n(t_m) \equiv n_{\rm max} \sim 4$.
Soon after that, the second term becomes dominant and $n(t)$ quickly decreases to zero.
We find that these background dynamics of $n(t)$ are well approximated by the following equations
\begin{equation}
n \simeq \begin{cases}
\dfrac{c_1}{M_{\rm Pl}\gamma N} \quad (\bar{\varphi} \gtrsim \bar{\varphi}_t)  \\
(n_{\rm max}+\alpha)\left( 1- \dfrac{N(t_m) - N}{\beta} \right) \quad (\bar{\varphi} \sim \bar{\varphi}_t)  \\
0 \quad (\bar{\varphi} \lesssim \bar{\varphi}_t) 
\end{cases} \ , \qquad \bar{\varphi}_t \equiv \dfrac{M_{\rm Pl}}{c_1}\ln\left(\dfrac{B_2}{B_1}\right) \ , \label{eq: na}
\end{equation}
where $\bar{\varphi}_t$ is the field value for which the first term becomes equal to the second term in \eqref{eq: I} and we introduced the fitting parameters $\alpha, \beta$ at the transition stage of $n(t)$.

Such a transition behavior affects the evolution of gauge field on super-horizon scales.
In Figure \ref{fig:deltaA}, we describe the time evolutions of the logarithmic energy density of electric field with several momentum scales until inflation ends.
$N_k(t_k)$ is the number of e-folding when the mode function $E_k$ exits the horizon: $k = a(t_k)H$. 
When $n(t)$ becomes greater than 2, the gauge field starts to grow in time and gets the maximum value at $t = t_{\rm peak}$.
After the transition occurs, it is not amplified any longer and decreases in time.
We analyze this behavior of the gauge field in appendix \ref{ap: a} and one finds that its description is far from the conventional analytical solution \eqref{eq: Hankel}.
Namely, the fluctuations on large scales (including the scale of CMB window) are not sufficiently enhanced by the gauge field.
Instead, a short-term amplification of gauge field takes place at the late stage of inflation
and we demonstrate that the copious gauge field production mostly occurs on the scales where fluctuations cross the horizon at $n(t) \sim 2 \ (N_k \sim 10)$.

%
%

\section{Generation of scalar mode}
\label{Perturbation Dynamics}

In this section, we analytically estimate the two-point correlation functions of scalar and tensor mode sourced by the gauge field.

\subsection{Primordial power spectrum}
Firstly, we calculate the perturbation of inflaton sourced by the gauge field.
The Fourier transformations of $\delta\varphi$ is as usual
\begin{align}
&\delta \varphi(t,\bm x)= 
\int \frac{\dd^3 k}{(2\pi)^3} e^{i\bm{k}\cdot\bm{x}}
\hat{\delta \varphi}_{\bm k}(t) \ .
\end{align}
Straightforwardly we get the EoMs for $\hat{\delta\varphi}$:
\begin{align}
&\left[\partial_x^2+1-\frac{2-\bar{V}_{\varphi\varphi}/H^2}{x^2}\right](a\hat{\delta\varphi}_{\bm{k}})
\simeq a^3\dfrac{2}{k^2}\dfrac{\bar{I}_\varphi}{\bar{I}}\hat{\delta\rho}_{E, \bm{k}} \ , \label{dphi EoM} \\
&\hat{\delta\rho}_{E, \bm{k}} = \int\dfrac{d\bm{p}}{(2\pi)^3}\dfrac{1}{2}\left(\hat{E}^X_{\bm{p}}e^X_{i}(\hat{\bm{p}}) + i\hat{E}^Y_{\bm{p}}e^Y_{i}(\hat{\bm{p}}) \right)\left(\hat{E}^X_{\bm{k}-\bm{p}}e^X_{i}(\widehat{\bm{k}-\bm{p}}) + i\hat{E}^Y_{\bm{k}-\bm{p}}e^Y_{i}(\widehat{\bm{k}-\bm{p}}) \right) \ , 
\end{align} 
where the mass term $\bar{V}_{\varphi\varphi}$ can be ignored due to the slow-roll suppression.
Note that we have neglected the contribution of magnetic field since it is more suppressed by the power of scale factor than that of electric field.
Then the solution of \eqref{dphi EoM} can be separated into two modes
\begin{equation}
\hat{\delta \varphi}_{\bm k} = \hat{\delta \varphi}_{\bm k, v} + \hat{\delta \varphi}_{\bm k, s} \ ,
\end{equation}
where $\hat{\delta \varphi}_{\bm k, v}$ is the usual vacuum fluctuation of $\delta\varphi$ satisfying the homogeneous solution whereas $\hat{\delta \varphi}_{\bm k, s}$ is the particular solution sourced by the gauge field.
The solution $\hat{\delta \varphi}_{\bm k, s}$ can be obtained by using the Green's function method
\begin{align}
a\hat{\delta\varphi}_{\bm{k},s}(x)
&= \dfrac{2}{k^2}\int\dd y\, G_{R}(x,y) a(y)^3\dfrac{\bar{I}_\varphi}{\bar{I}}\hat{\delta\rho}_E \ , \label{G function} \\
G_{R}(x,y) &\equiv -\Theta(y-x)\,(x^3-y^3)/(3xy) \ , \qquad y \equiv -k\tau' \ .
\end{align}
The retarded Green's function $G_{R}$ satisfies $\left[\partial_x^2-2/x^2\right]G_R (x,y)=\delta(x-y)$
where the gradient term and the mass term are ignored.

%
\begin{figure}[tbp]
\center
  \includegraphics[width=90mm]{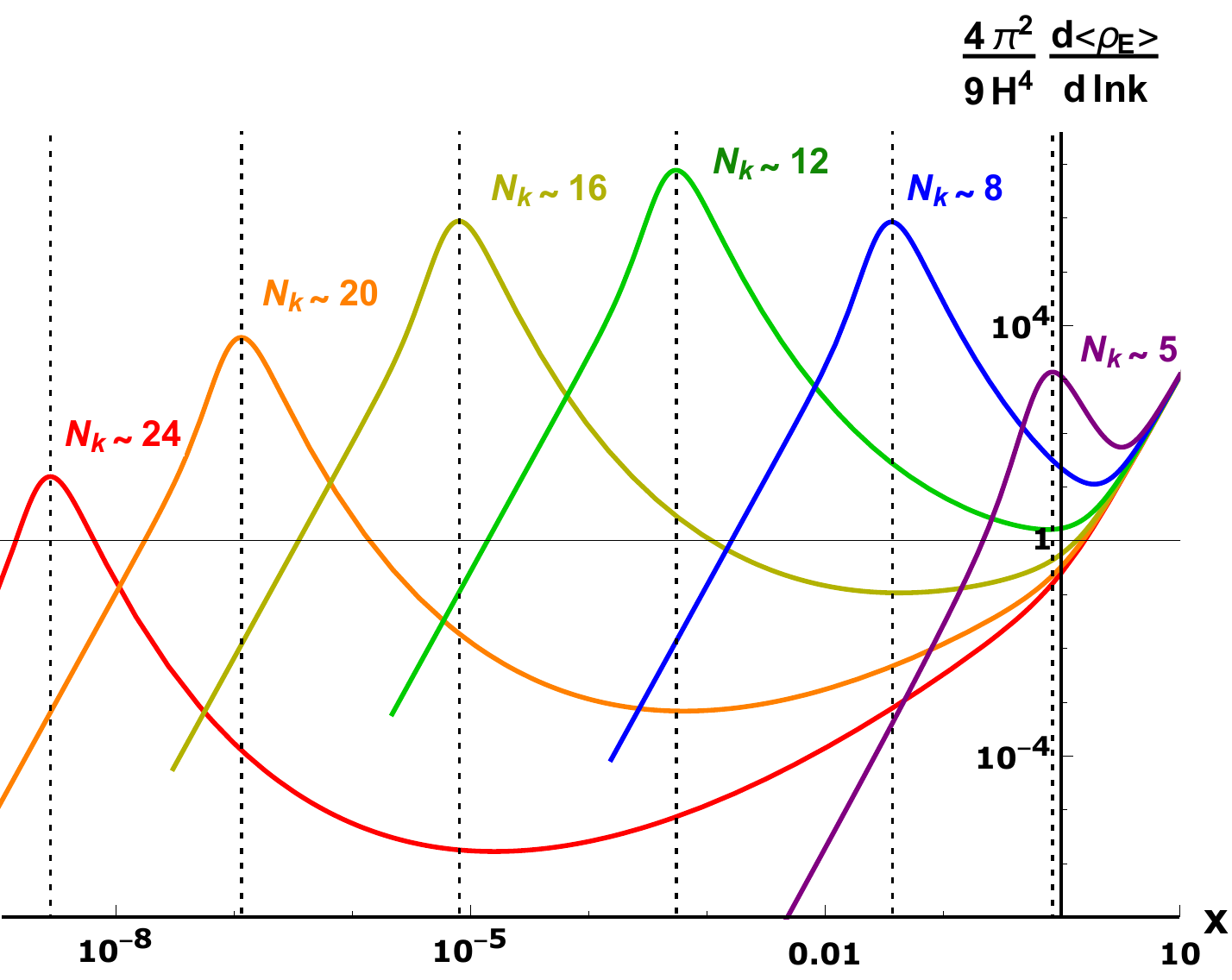}
  \caption
 {Time evolutions of the spectrum of electric energy density with several momentum scales exiting the horizon $x = 1$ from $N_k \sim 24$ (red) to $N_k \sim 5$ (purple). The time flows from right to left.
 The dotted line is at a time $t = t_{\rm peak}$ when the mode function gets its maximum value.
  We normalize the amplitude of electric density as that of mode function staying constant on the super-horizon scales.
  In this plot, we set the same values of model parameters as  in Figure \ref{fig:n}.}
 \label{fig:deltaA}
\end{figure}
%

Let us evaluate the two-point correlation function of inflaton perturbation $\langle \hat{\delta\varphi}_{\bm{k}} \hat{\delta\varphi}_{\bm{k}'} \rangle$.
Since the quantum operators of inflaton and gauge field are statistically independent, it is decomposed into the vacuum and the sourced power spectra
\begin{align}
\langle \hat{\delta\varphi}_{\bm{k}} \hat{\delta\varphi}_{\bm{k}'} \rangle &= \langle \hat{\delta\varphi}_{\bm{k},v} \hat{\delta\varphi}_{\bm{k}',v} \rangle + \langle \hat{\delta\varphi}_{\bm{k},s} \hat{\delta\varphi}_{\bm{k}',s} \rangle \notag \\
&\equiv (2\pi)^3\delta(\bm{k} + \bm{k}')\dfrac{2\pi^2}{k^3}\left(\mathcal{P}_{\delta\varphi,v}(k) + \mathcal{P}_{\delta\varphi,s}(k) \right) \ .
\end{align}
The power spectrum of vaccum mode is simply given by $\mathcal{P}_{\delta\varphi,v}(k) = H^2/(4\pi^2)|_{k=aH}$ evaluated at the time of horizon crossing.
On the other hand, $\mathcal{P}_{\delta\varphi,s}$ is the power spectrum that is sourced by the gauge field on the super-horizon scale.
Using the identities \eqref{eq: id1}-\eqref{eq: id3} in the appendix, we get
\begin{align}
\mathcal{P}_{\delta\varphi,s}(k)|_\tau &= \dfrac{k^3}{\pi^2H^4}\int\dfrac{d\bm{p}}{(2\pi)^3}\left[\cos^2(\theta_{\hat{\bm{p}}} + \theta_{\widehat{\bm{k}-\bm{p}}}) + 1\right]\left|\int_{\tau_{\rm min}}^{\tau}\dfrac{\dd \tau'}{\tau'}\, \dfrac{\bar{I}_\varphi}{\bar{I}}\dfrac{y^3-x^3}{3y^3} \, E_{p}E_{|\bm{k}-\bm{p}|} \right|^2 \ . \label{eq: calc1}
\end{align}
In the domain of time integration, we regard the lower limit of conformal time $\tau_{\rm min}$ as the time when both electric modes $E_{p}$ and $E_{|\bm{k}-\bm{p}|}$ have started their amplifications due to the tachyonic instability.
We neglect the contribution of energy density deep inside the horizon since it is a UV-divergent vacuum mode which needs to be renormalized away.
As can be seen from the previous section, the most contribution of time integration comes from the region where the gauge mode function is strongly peaked.
For our analytical convenience, we rewrite the mode function of electric field as
\begin{align}
aE_k &= H\bar I A_k\left(n-\dfrac{d\ln(\bar I A_k)}{d\ln x}\right) \label{eq: Elec}
\end{align}
and use the following Gaussian function well fitted around the peak of mode function
\begin{equation}
\bar{I} A_k(x) \simeq \bar{I} A^{\rm fit}_k(x) \equiv \dfrac{1}{\sqrt{2k}x}X_{\rm peak}(k)\exp{\left[-\dfrac{\left(\ln (\tau/\tau_{\rm peak})\right)^2}{\sigma^{2}}\right]} \label{eq: fit}
\end{equation}
(for the analytical estimate of $X_{\rm peak}(k)$ and $\sigma$, see the appendix \ref{ap: a}).
Using the relation $\bar{I}_{\varphi}/\bar{I} \simeq n/(M_{\rm Pl}\sqrt{2\epsilon_H})$ and the property of Gaussian support, we can evaluate \eqref{eq: calc1} as
\begin{align}
\mathcal{P}_{\delta\varphi,s}(k)|_{\tau\rightarrow 0} &\simeq \dfrac{H^4}{72\pi^2\Mpl^2\epsilon_H}\int\dfrac{d\bm{p}^{*}}{(2\pi)^3}\dfrac{\cos^2(\theta_{\hat{\bm{p}}} + \theta_{\widehat{\bm{k}-\bm{p}}}) + 1}{p^{*3}|\bm{k}-\bm{p}|^{*3}}X^2_{\rm peak}(p)X^2_{\rm peak}(|\bm{k}-\bm{p}|) \notag \\
&\times\left|\int_{\infty}^{-\infty}dw \, n\left(n+1+\dfrac{2w}{\sigma^{2}}\right)^2\exp{\left[-2\dfrac{w^2}{\sigma^{2}} \right]} \right|^2 \ , \label{eq: calc2}
\end{align}
where $\bm{p}^* \equiv \bm{p}/k, \ |\bm{k}-\bm{p}|^* \equiv |\bm{k}-\bm{p}|/k$.
Here we took the super-horizon limit $\tau\rightarrow 0$ and defined a new time variable $w \equiv \ln(\tau'/\tau_{\rm peak})$.
Since the integral receives its support almost around the peak $\tau = \tau_{\rm peak}$, the outer region of the time interval can be extended to infinity.
Performing the time integral, we approximately get
\begin{align}
\mathcal{P}_{\delta\varphi,s}(k)|_{\tau\rightarrow0}
&\simeq \dfrac{2\pi^2 \mathcal{F}^2}{9\Mpl^2\epsilon_H}\left(\dfrac{H}{2\pi}\right)^4\int\dfrac{d\bm{p}^{*}}{(2\pi)^3}\dfrac{\cos^2(\theta_{\hat{\bm{p}}} + \theta_{\widehat{\bm{k}-\bm{p}}}) + 1}{p^{*3}|\bm{k}-\bm{p}|^{*3}}X^2_{\rm peak}(p)X^2_{\rm peak}(|\bm{k}-\bm{p}|) \ , \label{eq: calc3}
\end{align}
where $\mathcal{F} = \mathcal{O}(30)$ is a numerical factor obtained by the time integration.
As is the momentum integral, the IR cutoff should be restricted to the regime where the gauge modes were inside the horizon at the start of inflation
\begin{equation}
p \ , |\bm{k} - \bm{p}| > \dfrac{1}{-\tau_{\rm in}} \ .
\end{equation}
It might seem that the momentum integral has most of its support at the two logarithmic poles $\bm{p}\rightarrow \bm{0}, \ \bm{k}$.
However, the gauge modes on large scales are not significantly amplified and do not contribute to the integration in our model.
Therefore, we can naively define the effective lower (and upper) momentum limit as the scale where $X_{\rm peak}$ becomes sufficiently smaller than its maximum value.
%
%
%
%
%
%

\begin{figure}[tbp]
\center
  \includegraphics[width=110mm]{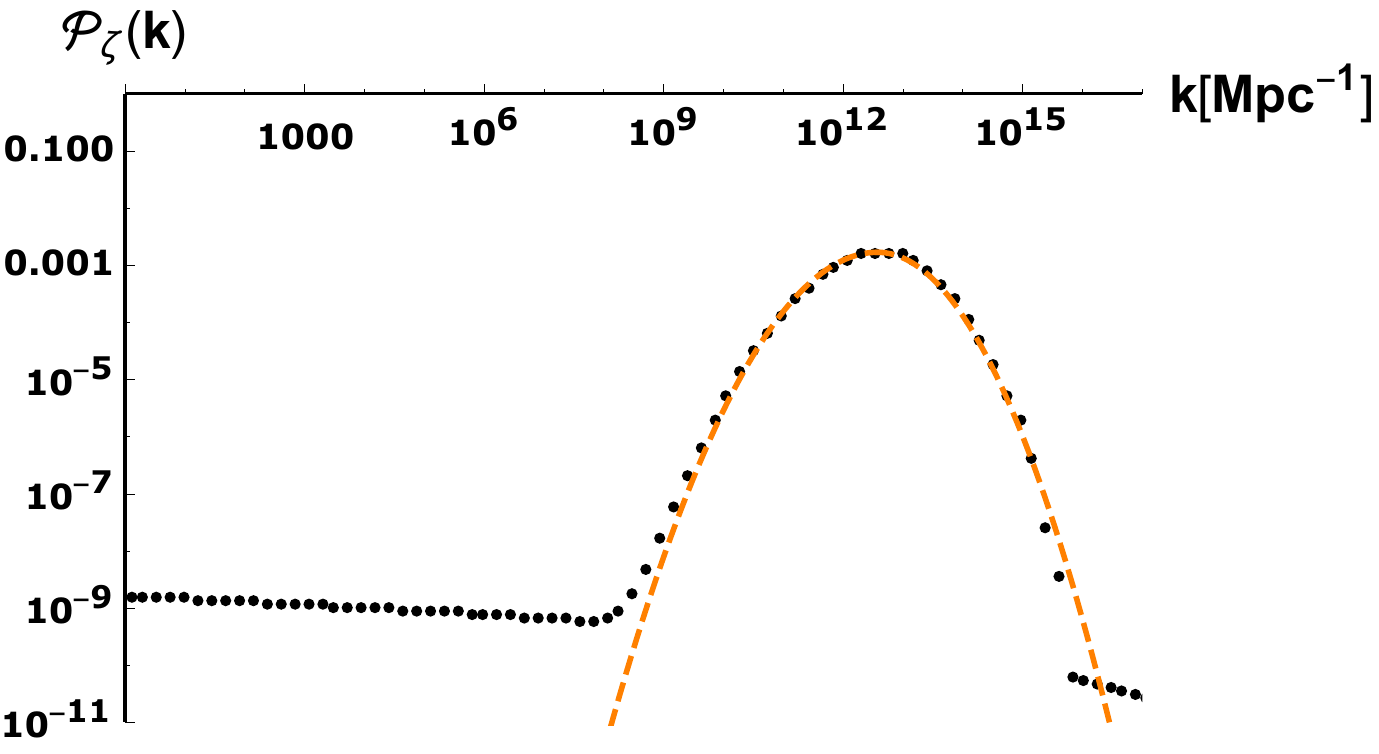}
  \caption
 {Plots of power spectrum of curvature perturbation with respect to the wave number $k [\text{Mpc}^{-1}]$.
 The black dots represent the numerical estimation of \eqref{eq: calc3}. 
 The orange dashed line shows the approximate fitting function \eqref{eq: gaus}.
 In this plot we set $B_2/B_1\simeq 8.8\times10^{25}$ and $c_1\simeq23$.}
 \label{fig:zeta}
\end{figure}

We numerically integrate \eqref{eq: calc3} by using the fitting function of $X_{\rm peak}$ \eqref{eq: fit2} and plot the power spectrum of curvature perturbation at flat-slicing
\begin{equation}
\zeta \equiv -\dfrac{H}{\dot{\bar{\varphi}}}\delta\varphi 
\end{equation}
in Figure \ref{fig:zeta}.
In this figure, we related the wave number $k$ of mode function to the number of e-foldings $N$ as
\begin{equation}
k = k_{\rm CMB}\exp\left( N_{\rm CMB} - N \right) \ ,
\end{equation}
where $k_{\rm CMB} = 0.002 ~\text{Mpc}^{-1}$ is the Planck pivot scale.
%
%
The spectral shape is bumped at around the scales crossing the horizon when $n(t) \sim 2$.
For our phenomenological predictions, we fit the following Gauss function to the bump of power spectrum
\begin{equation}
\mathcal{P}_{\zeta}(k) \simeq \mathcal{P}_{\zeta, v}(k) + A \exp\left[-\dfrac{(\ln(k/k_p))^2}{\sigma_\zeta^2}\right] \ . \label{eq: gaus}
\end{equation}
The first term $\mathcal{P}_{\zeta, v}$ is a power spectrum of vacuum mode and we normalize it on the CMB scales as $\mathcal{P}_{\zeta, v}(k_{\rm{CMB}}) \simeq 2.1\times10^{-9}$.
The second term characterizes the bump of the power spectrum and we set
\begin{equation}
A \simeq 1.7\times10^{-3} \ , \qquad k_p = 3.8\times 10^{12}\text{Mpc}^{-1} \ , \qquad \sigma^2_\zeta \simeq 2.4^2\Theta(k_p-k) + 2.1^2\Theta(k - k_p)
\end{equation}
in the plot of Figure \ref{fig:zeta}.
We use \eqref{eq: gaus} in the computation of PBH mass spectrum and secondary gravitational waves discussed in the next section.

%
%

\subsection{Mass spectrum of PBHs}

Finally, we estimate the mass spectrum of PBHs in our model.
There have been extensive studies for the calculation of the PBH abundance depending on the methods of calculations or some uncertainties \cite{PinaAvelino:2005rm, Bugaev:2011wy, Byrnes:2012yx, Young:2013oia, Green:2004wb, Yoo:2018esr}.
In this paper, we follow a simple method based on the Press-Schechter formalism because our model has enough parameter space to adjust the required value of the power spectrum to produce the PBHs as dark matter.

Since the PBH formation occurs at the horizon crossing, we can relate the scale of perturbation with the PBH mass \cite{Inomata:2017vxo}
\begin{align}
M(k) &\equiv \left.\tilde{\gamma}\rho\dfrac{4\pi H^{-3}}{3}\right|_{k=aH} \notag \\
&\simeq 10^{20}\text{g}\left(\dfrac{\tilde{\gamma}}{0.2}\right)\left(\dfrac{g_*}{106.75}\right)^{-\tfrac{1}{6}}\left(\dfrac{k}{7\times10^{12}\text{Mpc}^{-1}}\right)^{-2} \ ,
\end{align}
where $\tilde{\gamma}$ is a numerical factor depending on the gravitational collapse and $g_*$ is an effective number of relativistic degrees of freedom at PBH formation.
Here we use the simple analytic estimation $\tilde{\gamma}\simeq 0.2$ \cite{Carr:1975qj}.
Based on the Press-€"Schechter formalism, the PBH formation rate is obtained by estimating the averaged density perturbation over the horizon size $R$:
\begin{align}
	\bar\delta_R (\boldsymbol x, R)
	&=\int \df^3 yW(|\boldsymbol x-\boldsymbol y|,R)\delta(\boldsymbol y)
	= \int \frac{\df^3 k}{(2\pi)^3}
	\tilde W(kR)\delta_k e^{i\boldsymbol k\cdot \boldsymbol x } \ ,
\end{align}
where $W(\boldsymbol x,R)$ and $\tilde W(kR)$ are window functions in position and momentum space.
In this work, we adopt the two types of window function; Gaussian type and real-space top-hat type
\begin{align}
\tilde W(kR) = \begin{cases}
\exp\left(-\dfrac{(kR)^2}{2}\right) \qquad (\text{Gaussian}) \\
3\left(\dfrac{\sin(kR)-kR\cos(kR)}{(kR)^3}\right) \qquad (\text{Top-hat})
\end{cases} \ ,
\end{align}
whose choice results in a large uncertainty on the PBH abundance \cite{Ando:2018qdb}.
The top-hat window function has been motivated by the numerical study of PBH formation, where the Compaction function is used as the definition of overdensity \cite{Shibata:1999zs}.
On the other hand, the Gaussian window function has been used for the conservative estimation of the PBH abundance which predicts an inefficient PBH formation and hence requires large perturbations to explain PBHs as dark matter compared to the top-hat window function.
As the conservative method, when we use the Gaussian type, we simply neglect the transfer function and the effect of nonlinear relationship between $\bar \delta_k$ and $\zeta_k$ as we will discuss below.

Let us evaluate the probability distribution function of $\bar{\delta}_k$ by $\mathcal{P}_\zeta(k)$.
In Press-Schechter formalism, PBHs are formed when $\bar{\delta}_R(\bm{x}, R)$ overcomes the threshold value $\delta_c$.
In this paper, as the typical threshold value, we adopt $\delta_c=0.53$ in the comoving gauge given by the numerical simulations \cite{Shibata:1999zs, Harada:2015yda, Yoo:2018esr}.
It is also known that the nonlinear relation between $\bar \delta_k$ and $\zeta_k$ suppresses the PBH formation rate and requires the power spectrum about a factor 2 larger than in the case of linear estimation \cite{Kawasaki:2019mbl,Young:2019yug,DeLuca:2019qsy}.
The non-linear effect has been calculated based on the definition of Compaction function which results in the analysis of the top-hat window function.
For the case of Gaussian window function, however, the calculation of nonlinearity is unclear and we simply disregard its effect.
Therefore, for the case of top-hat type we use the linear relationship $\delta_k=\frac{4}{9}\left( \frac{k}{aH} \right)^2 \zeta_k$ and include the nonlinear effect by the effective threshold value $\delta_c \rightarrow \alpha_{\rm{NL}}\delta_c$ with $\alpha_{\rm{NL}} = \sqrt{2}$.
The variance of $\bar\delta_R (\boldsymbol x)$ is given by
\begin{align}
	\sigma_\delta(R)^2
	= \int \df (\ln p)
	\frac{16}{81}
	|\tilde W(pR)|^2|T(pR)|^2
	\left( pR \right)^4 \mathcal P_\zeta(p) \ , \label{eq: sig}
\end{align}
where the transfer function $T(pR)$ describes the time evolution of the perturbations in subhorizon \cite{Scott}
\begin{equation}
T(x) = 3\dfrac{\sin(x/\sqrt{3}) - (x/\sqrt{3})\cos(x/\sqrt{3})}{(x/\sqrt{3})^3}
\end{equation}
which is used together with the top-hat window function in order to suppress the contribution of higher momentum modes.
Since the curvature perturbation is the quadratic function of the gauge field, we assume that the averaged density perturbation obeys the $\chi^2$-distribution: $\bar \delta_R=g^2-\sigma_g^2$, where $g$ follows the Gaussian distribution with the variance $\sigma_g$.
In this case, the probability distribution function of $\bar \delta_R$ is given by \cite{Linde:2012bt, Lyth:2012yp, Byrnes:2012yx, Bugaev:2011wy}
\begin{align}
	P_\delta(\bar \delta_R)\df \bar\delta_R
		&= \frac{1}{\sqrt{2\pi \sigma_g^2(\bar\delta_R+\sigma_g^2)}}
	 \exp\left( - \frac{\bar \delta_R+\sigma_g^2}{2\sigma_g^2} \right)
	 \df \bar\delta_R \ ,
\end{align}
where $\sigma_\delta^2= 2\sigma_g^4$.
Suppose that the PBH formation rate is 
\begin{align}
	\beta(M)
	=\int_{\alpha_{\text{NL}}\delta_c} P_\delta(\bar \delta_R)\df \bar\delta_R \bigg|_{R=k^{-1}(M)}
	=\text{Erfc}\left(  \sqrt{\frac{1}{2}+\frac{\alpha_{\text{NL}}\delta_c}{\sigma_\delta(R)\sqrt{2}}}\right) \ ,
\end{align}
the mass spectrum of the PBH is given by
\begin{align}
f_{\text{PBH}}(M) &\equiv \dfrac{\Omega_{\text{PBH}}(M)}{\Omega_c} \notag \\
&\simeq \left(\dfrac{\beta(M)}{8.0\times10^{-15}} \right)\left(\dfrac{0.12}{\Omega_ch^2}\right)\left(\dfrac{\tilde{\gamma}}{0.2}\right)^{\tfrac{3}{2}}\left(\dfrac{106.75}{g_*(T_M)}\right)^{\tfrac{1}{4}}\left(\dfrac{M}{10^{20}\text{g}}\right)^{-\tfrac{1}{2}} \ .
\end{align}

\begin{figure}[htbp]
\begin{center}
\includegraphics[width=0.6\textwidth]{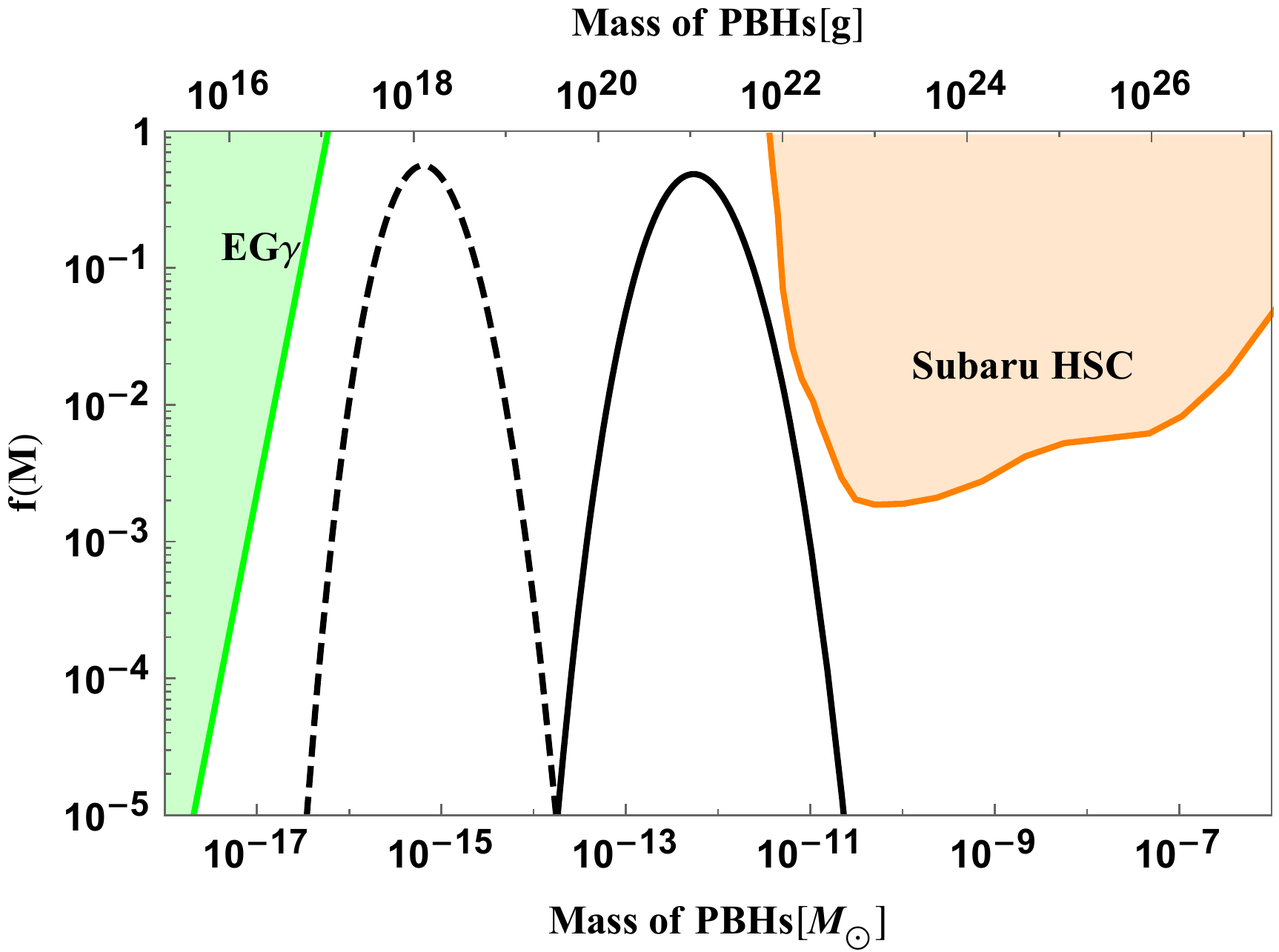}
\end{center}
\caption{
Plot of mass spectrum of PBHs in our model with $\{B_2/B_1\simeq8.8\times10^{25}, \ c_1\simeq23\}$ (solid line) and $\{B_2/B_1\simeq 2.6\times10^{15}, \ c_1\simeq17\}$ (dashed line) for the case of top-hat window function.
The green region around $M<\mathcal O(10^{17}) \text g$ is excluded by the extragalactic gamma ray \cite{Carr:2009jm} and the orange region  $\mathcal O(10^{22}) \text g <M $ is excluded by the Subaru/HSC \cite{Niikura:2017zjd}.
}
\label{fig:PBHspectrum}
\end{figure}

We use the fitting function of $\mathcal{P}_\zeta$ \eqref{eq: gaus} in the computation of \eqref{eq: sig} and plot the mass spectrum of PBHs with $10^{16} \text{g}< M < 10^{23} \text{g}$ ~in Figure \ref{fig:PBHspectrum}.
For the PBH mass region around here, we need to take into account the observational constraints on the PBH dark matter given by the micro-lensing events and the Hawking radiation.
Since light PBHs evaporate by the Hawking radiation and decay into photons, the observations of extragalactic gamma-ray constrain the abundance of PBH with mass lighter than $\mathcal O(10^{17}) \text{g}$ \cite{Carr:2009jm}.
On the other hand, heavy PBHs can be detected by the effect of the gravitational lenses. The Subaru/HSC constrains the PBHs with heavier mass than $\mathcal O(10^{22}) \text{g}$ \cite{Niikura:2017zjd}. 
Although there have been other constraints discussed in the mass window $10^{17} \text{g} \lesssim M \lesssim 10^{22} \text{g}$ ~such as the GRB femtolensing events \cite{Barnacka:2012bm}, the dynamical capture of PBHs by stars \cite{Capela:2012jz, Capela:2013yf, Pani:2014rca, Capela:2014ita}, the ignition of white dwarfs by PBHs \cite{Graham:2015apa}, the recent studies have revisited them and claimed that this mass window still opens for the PBH as dark matter \cite{Katz:2018zrn, Montero-Camacho:2019jte}.
In Figure \ref{fig:PBHspectrum}, we use two sets of background parameter values and demonstrate two PBH mass spectra getting peaked at different mass scales (solid and dashed lines).
Both mass spectra can explain all of the dark matter avoiding the current constraints on the PBH abundance.
As we will see in the next section, these mass spectra predict the gravitational wave power spectra enhanced at different frequencies, which are testable with the future space-based laser interferometers.

\section{Generation of tensor modes}
\label{tensor}

In this section, we calculate the power spectrum of tensor mode sourced by the gauge field.
We firstly discuss the primordial tensor power spectrum.
Subsequently, we compute the induced gravitational waves sourced by the scalar field at second order after inflation.

\subsection{Primordial power spectrum}
\label{Primordial}

The tensor perturbation is given by the fluctuations of the spacial metric component $g_{ij}(t, \bm{x}) = a(t)^2(\delta_{ij} + \tfrac{1}{2}h_{ij}(t,\bm{x}))$ which obeys the following equation of motion at leading order
\begin{equation}
\left[ \partial_t^2+3H\partial_t - \dfrac{\nabla^2}{a^2} \right]h_{ij} \simeq -\dfrac{4}{\Mpl^2}\Pi^{lm}_{ij}E_lE_m \ ,
\end{equation}
where $\Pi^{lm}_{ij}$ is the transverse-traceless projector defined by
\begin{equation}
\Pi^{lm}_{ij} \equiv \Pi^l_i\Pi^m_j - \dfrac{1}{2}\Pi_{ij}\Pi^{lm} \ , \qquad \Pi_{ij} \equiv \delta_{ij} - \dfrac{\partial_i\partial_j}{\nabla^2} \ .
\end{equation}
We decompose $h_{ij}$ into the linear polarization tensors $e^{s}_{ij}(\hat{\bm k})  \ (s = +,\times)$ (these definitions are given in appendix \ref{ap: p}) in Fourier space
\begin{align}
h_{ij}(t,\bm{x}) &= \int\dfrac{d\bm{k}}{(2\pi)^3}\hat{h}_{ij}(\bm{k}, t)e^{i\bm{k}\cdot\bm{x}} \notag \\
&= \int\dfrac{d\bm{k}}{(2\pi)^3}\left[e^+_{ij}(\hat{\bm{k}})\hat{h}^+_{\bm
{k}}(t) + e^\times_{ij}(\hat{\bm{k}})\hat{h}^\times_{\bm
{k}}(t)\right]e^{i\bm{k}\cdot\bm{x}} \ .
\end{align}
Using the following relations $\hat{h}^s_{\bm
{k}} = e^{s}_{ij}(\hat{\bm k})\hat{h}_{ij}(\hat{\bm k})$ and $\Pi_{ij}^{lm}e^s_{lm}(\hat{\bm k}) = e^s_{ij}(\hat{\bm k})$,
we obtain
\begin{align}
&\left[ \partial_x^2 +1 - \dfrac{2}{x^2} \right](a\hat h^{s}_{\bm k}) = -e^{s}_{ij}(\hat{\bm{k}})\dfrac{4a^3}{k^2\Mpl^2} \notag \\
&\times\int\dfrac{d\bm p}{(2\pi)^3}\left(\hat{E}^X_{\bm{p}}e^X_{i}(\hat{\bm{p}}) + i\hat{E}^Y_{\bm{p}}e^Y_{i}(\hat{\bm{p}}) \right)\left(\hat{E}^X_{\bm{k}-\bm{p}}e^X_{j}(\widehat{\bm{k}-\bm{p}}) + i\hat{E}^Y_{\bm{k}-\bm{p}}e^Y_{j}(\widehat{\bm{k}-\bm{p}}) \right) \ . \label{eq: ten}
\end{align}
Then we get two solutions of \eqref{eq: ten}:
$\hat{h}^s_{\bm k} = \hat{h}^s_{\bm k, \rm v} + \hat{h}^s_{\bm k, \rm s}$.
Using the identities of polarization tensors \eqref{eq: id4}-\eqref{eq: id7} in the appendix, the sourced tensor modes are represented by
\begin{align}
a\hat{h}^+_{\bm{k},s} &= \dfrac{2\sqrt{2}}{k^2\Mpl^2}\int d y ~a^3G_R(x, \ y)\int\dfrac{d\bm p}{(2\pi)^3}~\left(\hat{E}^X_{\bm{p}}\hat{E}^X_{\bm{k}-\bm{p}}\cos\theta_{\hat{\bm p}}\cos\theta_{\widehat{\bm k - \bm p}} + \hat{E}^Y_{\bm{p}}\hat{E}^Y_{\bm{k}-\bm{p}} \right) \ , \\
a\hat{h}^\times_{\bm{k},s} &= -\dfrac{2\sqrt{2}}{k^2\Mpl^2}\int d y ~a^3G_R(x, \ y)\int\dfrac{d\bm p}{(2\pi)^3}~\left( \hat{E}^X_{\bm{p}}\hat{E}^Y_{\bm{k}-\bm{p}}\cos\theta_{\hat{\bm p}} + \hat{E}^Y_{\bm{p}}\hat{E}^X_{\bm{k}-\bm{p}}\cos\theta_{\widehat{\bm k - \bm p}} \right) \ .
\end{align}
Therefore defining the dimensionless power spectrum of tensor modes
\begin{align}
\langle \hat{h}^s_{\bm{k}} \hat{h}^{s'}_{\bm{k}'} \rangle &= \langle \hat{h}^s_{\bm{k},\rm v} \hat{h}^{s'}_{\bm{k}',\rm v} \rangle + \langle \hat{h}^s_{\bm{k},\rm s} \hat{h}^{s'}_{\bm{k}',\rm s} \rangle \notag \\
&\equiv (2\pi)^3\delta^{ss'}\delta(\bm{k} + \bm{k}')\dfrac{2\pi^2}{k^3}\left(\mathcal{P}_{h,\rm v}(k) + \mathcal{P}^{ss}_{h, \rm s}(k) \right) \ ,
\end{align}
one can find
\begin{align}
\mathcal{P}^{++}_{h, \rm s}(k)|_{\tau} &= \dfrac{8k^3}{\pi^2H^4\Mpl^4}\int\dfrac{d\bm{p}}{(2\pi)^3}\left[ \cos^2\theta_{\hat{\bm p}}\cos^2\theta_{\widehat{\bm k - \bm p}} + 1\right]\left|\int_{\tau_{\rm min}}^\tau \dfrac{d\tau'}{\tau'} \dfrac{x^3 - y^3}{3y^3}E_pE_{|\bm k -\bm p|}\right|^2 \ , \\
\mathcal{P}^{\times\times}_{h, \rm s}(k)|_{\tau} &= \dfrac{8k^3}{\pi^2H^4\Mpl^4}\int\dfrac{d\bm{p}}{(2\pi)^3}\left[ \cos^2\theta_{\hat{\bm p}} + \cos^2\theta_{\widehat{\bm k - \bm p}} \right]\left|\int_{\tau_{\rm min}}^\tau \dfrac{d\tau'}{\tau'} \dfrac{x^3 - y^3}{3y^3}E_p E_{|\bm k -\bm p|}\right|^2 \ .
\end{align}
As is a similar way to the previous section, we perform the integral by using the fitting function of the gauge mode function
\begin{align}
\mathcal{P}^{++}_{h, \rm s}(k)|_{\tau\rightarrow0} &\simeq \dfrac{2}{\pi^2} \dfrac{\mathcal{G}^2H^4}{9\Mpl^4}\int\dfrac{d\bm{p}^*}{(2\pi)^3}\left[\cos^2\theta_{\hat{\bm p}}\cos^2\theta_{\widehat{\bm k - \bm p}} + 1\right]\dfrac{X^2_{\rm peak}(p)X^2_{\rm peak}(|\bm{k}-\bm{p}|)}{p^{*3}|\bm k - \bm p|^{*3}} \ , \\
\mathcal{P}^{\times\times}_{h, \rm s}(k)|_{\tau\rightarrow0} &\simeq \dfrac{2}{\pi^2}\dfrac{\mathcal{G}^2H^4}{9\Mpl^4}\int\dfrac{d\bm{p}^*}{(2\pi)^3}\left[ \cos^2\theta_{\hat{\bm p}} + \cos^2\theta_{\widehat{\bm k - \bm p}} \right]\dfrac{X^2_{\rm peak}(p)X^2_{\rm peak}(|\bm{k}-\bm{p}|)}{p^{*3}|\bm k - \bm p|^{*3}} \ ,
\end{align}
where $\mathcal{G} = \mathcal{O}(10)$ is a numerical factor obtained by the time integration.
%
%
Therefore, the resultant tensor power spectrum is similar in shape to that of curvature perturbation.
However, the enhancement ratio of tensor mode $\mathcal{R}_h \equiv \mathcal{P}_{h, \rm s}/\mathcal{P}_{h, \rm v}$ is smaller than that of scalar mode $\mathcal{R}_\zeta \equiv \mathcal{P}_{\zeta, \rm s}/\mathcal{P}_{\zeta, \rm v}$ by a factor of tensor-to-scalar ratio
\begin{equation}
\dfrac{\mathcal{R}_h}{\mathcal{R}_\zeta} \sim \dfrac{\mathcal{G}^2}{16\mathcal{F}^2}r_{\rm v} \ , \qquad r_{\rm v} \equiv 16\epsilon_H \ . \label{eq: ratio}
\end{equation}

\subsection{Induced power spectrum}
\label{Induced}

Next, we calculate the tensor mode induced by the second-order scalar modes after inflation, following the previous method \cite{Inomata:2016rbd,Ando:2017veq,Ando:2018qdb}.
We take the conformal Newtonian gauge 
\begin{equation}
ds^2 = a(\tau)^2\left[ -(1+2\Phi)d\tau^2 + \left\{(1-2\Psi)\delta_{ij} + \dfrac{1}{2}h_{ij}\right\}dx^idx^j\right] \label{eq: re} \ ,
\end{equation}
where we neglected vector perturbations.
We assume that the two scalar perturbations $\Phi$ and $\Psi$ satisfy the condition of no anisotropic pressure: $\Phi = \Psi$.
We are interested in the induced tensor modes which enter the horizon at the radiation-dominated era $\tau < \tau_{\rm eq}$.
Then the equation of motion of tensor mode is given by
\begin{align}
&\left[\partial_\tau^2  - \nabla^2 \right](ah_{ij}) = -4a\Pi_{ij}^{lm}\mathcal{S}_{lm} \ , \\
&\mathcal{S}_{ij} \equiv 4\Psi\partial_i\partial_j\Psi + 2\partial_i\Psi\partial_j\Psi - \dfrac{1}{\mathcal{H}^2}\partial_i(\Psi' + \mathcal{H}\Psi)\partial_j(\Psi' + \mathcal{H}\Psi) \ .
\end{align}
Then the solution of induced tensor mode in momentum space is given by
\begin{align}
h^s_{\bm k,\rm i}(\tau) &= \dfrac{4}{a(\tau)}\int_0^\infty d\tau' a(\tau')G_k(\tau, \tau')\mathcal{S}_{\bm k}(\tau') \ , \\
G_k(\tau, \tau') &\equiv \Theta(\tau - \tau')\dfrac{1}{k}\sin(k\tau-k\tau') \ , \\
\mathcal{S}_{\bm k}(\tau) &= e_{ij}^{s}(\hat{\bm{k}})\int\dfrac{d\bm{p}p_ip_j}{(2\pi)^3}\left[ 3\Psi_{\bm{p}}\Psi_{\bm k - \bm{p}} + \dfrac{1}{\mathcal{H}}\left(\Psi_{\bm{p}}\Psi'_{\bm k - \bm{p}} + \Psi'_{\bm{p}}\Psi_{\bm k - \bm{p}} \right) + \dfrac{1}{\mathcal{H}^2}\Psi'_{\bm{p}}\Psi'_{\bm k - \bm{p}}  \right] \label{eq: Sk} \ .
\end{align}
In order to evaluate \eqref{eq: Sk}, we decompose $\Psi_{\bm k}(\tau)$ into the primordial field $\psi_{\bm k}$ and the transfer function $\Psi(k\tau)$:
\begin{align}
\Psi_{\bm k}(\tau) &= \psi_{\bm k}\Psi(k\tau) \ , \\
\Psi(k\tau) &= \dfrac{9}{(k\tau)^2}\left[ \dfrac{\sin(k\tau/\sqrt{3})}{k\tau/\sqrt{3}} - \cos(k\tau/\sqrt{3}) \right] \ .
\end{align}
Using $\psi_{\bm k} = -2\zeta_{\bm k}/3$ and $a(\tau) = a_0\tau/\tau_0$ at the radiation-dominated era, the resultant spectrum is given by
\begin{align}
\langle h^s_{\bm k, \rm i}(\tau)h^{s'}_{\bm k', \rm i}(\tau) \rangle &= \dfrac{128}{81}\left(\dfrac{a_0}{a\tau_0}\right)^2\dfrac{1}{k^3k'^3}\int\dfrac{d\bm{p}d\bm{q}}{(2\pi)^6}e^s_{ij}(\hat{\bm k})p_ip_je^{s'}_{kl}(\hat{\bm k}')q_kq_l \notag \\
&\times\mathcal{I}(p/k,|\bm k - \bm p|/k, k\tau)\mathcal{I}(q/k',|\bm k' - \bm q|/k', k'\tau)\langle \zeta_{\bm p}\zeta_{\bm k - \bm p}\zeta_{\bm q}\zeta_{\bm k' - \bm q} \rangle \ , \label{eq: ind}
\end{align}
where $\mathcal{I}$ is given by
\begin{align}
\mathcal{I}(\nu, u, x) &= \int_0^xdy y\sin(x-y)\left[ 3\Psi(\nu y)\Psi(u y) + y\{ \Psi(\nu y)u\dfrac{d\Psi(u y)}{d(uy)} + \nu\dfrac{d\Psi(\nu y)}{d(\nu y)}\Psi(u y) \} \right. \notag \\
&\left. + y^2u \nu \dfrac{d\Psi(\nu y)}{d(\nu y)}\dfrac{d\Psi(u y)}{d(uy)} \right]
\end{align}
in terms of new variables $\nu \equiv p/k$ and $u \equiv |\bm k - \bm p|/k$.
Note that we have re-defined the dimensionless time variables as $x \equiv k\tau, \ y \equiv k\tau'$.

\begin{figure}[htbp]
 \begin{minipage}{0.3\hsize}
  \begin{center}
   \includegraphics[width=30mm]{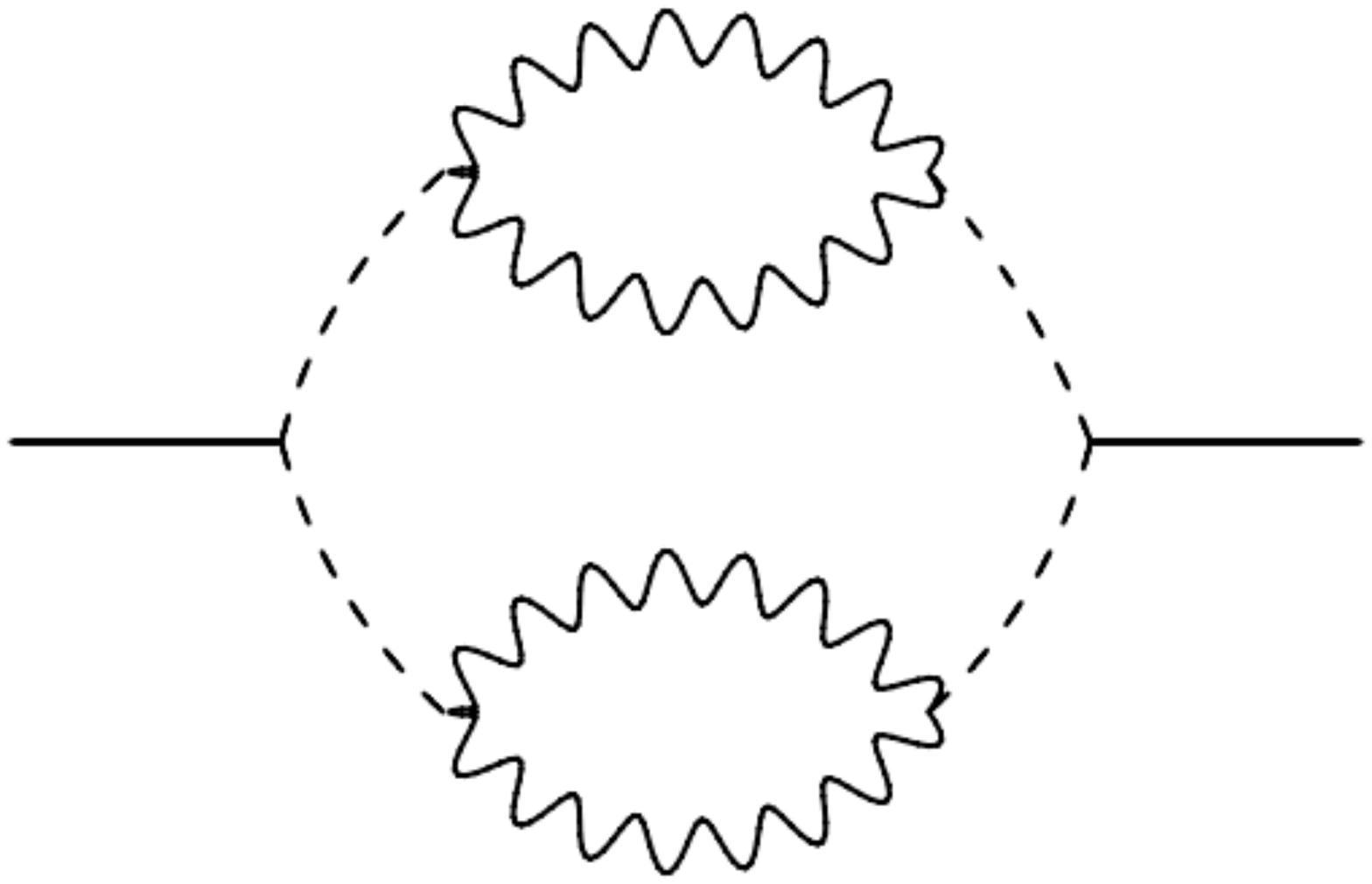}
  \end{center}
 \end{minipage}
 \begin{minipage}{0.3\hsize}
 \begin{center}
  \includegraphics[width=30mm]{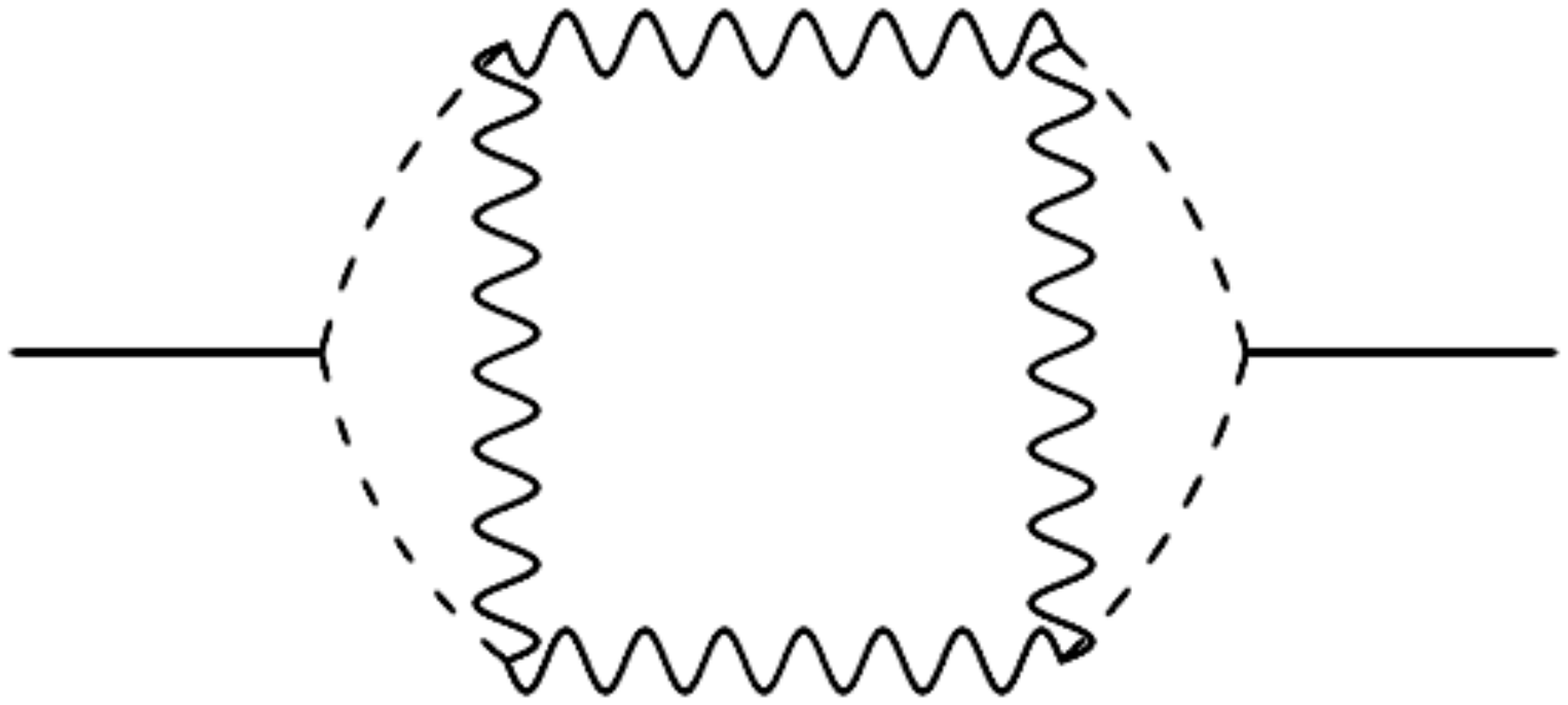}
 \end{center}
 \end{minipage}
 \begin{minipage}{0.3\hsize}
 \begin{center}
  \includegraphics[width=30mm]{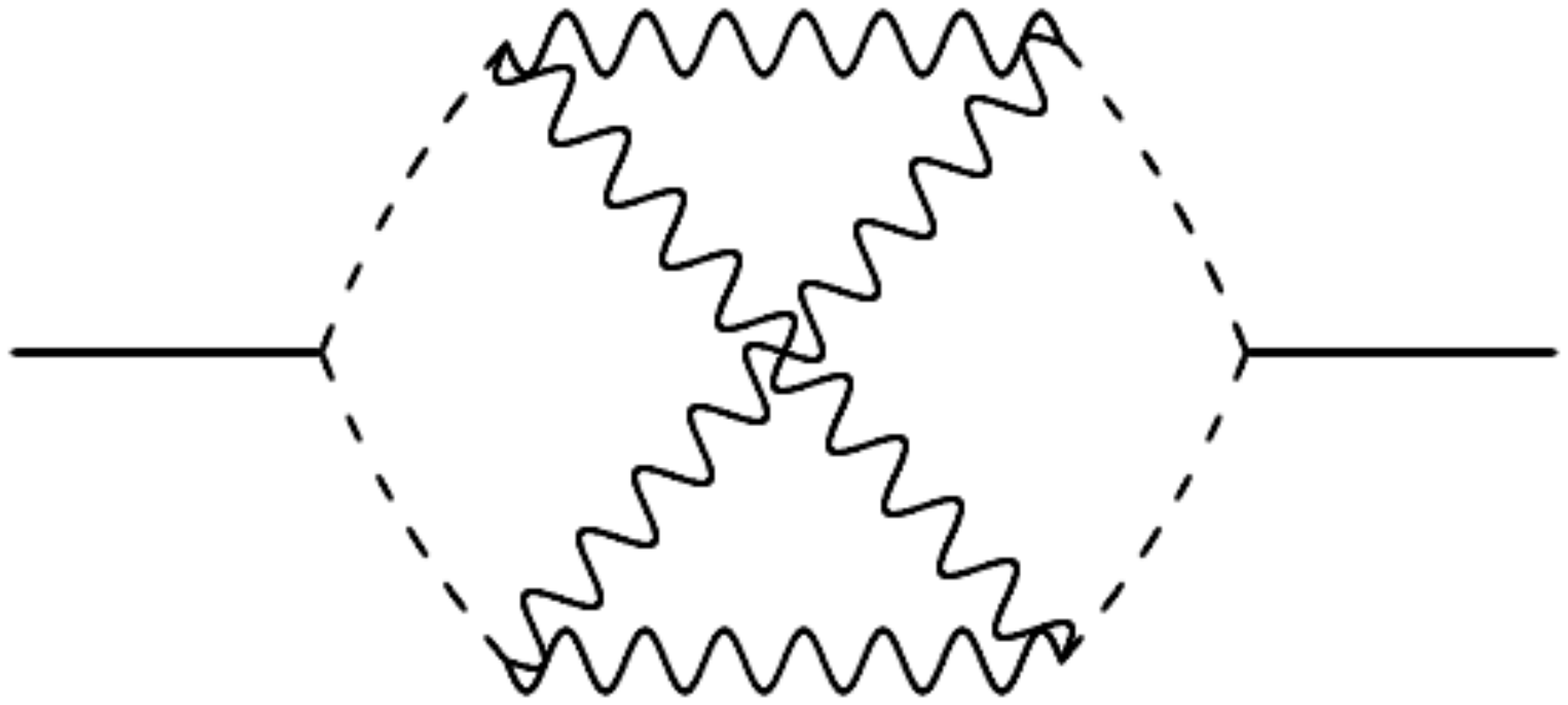}
 \end{center}
  \end{minipage}
 \caption{Loop contributions to the power spectrum of induced gravitational waves. We label these diagrams as ``Reducible" (left), ``Planar" (center) and ``Non-Planar" (right). The external solid line represents the gravitational wave perturbation $h^{+/\times}$. The intermediate dashed (wiggly) line represents the curvature perturbation $\zeta$ (the gauge field $A$).}
  \label{fig:diagrams}
\end{figure}

It should be mentioned that the 4-point correlation function $\langle\zeta^4\rangle$ in \eqref{eq: ind} cannot be completely replaced with the production of a 2-point correlation function $\langle\zeta^2\rangle$ because the main contribution of $\zeta$ is composed of the quadratic operators of amplified gauge field $\zeta \sim AA$.
At this time, there appear three kinds of loop diagrams in the computation of \eqref{eq: ind} (see Figure \ref{fig:diagrams}).
The left diagram in Figure \ref{fig:diagrams} is expressed as a three 1-loop diagram, which can be reduced to the production of the sourced power spectrum of curvature perturbation $\mathcal{P}_\zeta$.
Following the previous studies \cite{Garcia-Bellido:2017aan, Unal:2018yaa}, we call this diagram as ``Reducible" diagram.
The other two contributions are expressed as 3-loop diagrams which cannot be further factorized (``Planar" and ``Non-Planar" diagrams).
It has been found that Reducible diagram and Planar diagram have the same order contributions to the spectrum while that from Non-Planar diagram is suppressed compared to the other two diagrams \cite{Garcia-Bellido:2017aan}.
Although there will be a difference of spectral shapes between the resultant spectra from these diagrams, in this work we only take into account the contribution from Reducible diagram.
Therefore, by taking some symmetries we get
\begin{align}
\langle h^s_{\bm k, \rm i}(\tau)h^{s'}_{\bm k', \rm i}(\tau) \rangle &= \dfrac{256}{81}\left(\dfrac{a_0}{a\tau_0}\right)^2\dfrac{1}{k^6}\int\dfrac{d\bm{p}}{(2\pi)^3}e^s_{ij}(\hat{\bm k})p_ip_je^{s'*}_{kl}(\hat{\bm k})p_kp_l \dfrac{2\pi^2}{p^3}\dfrac{2\pi^2}{|\bm{k}-\bm{p}|^3} \notag \\
&\times\mathcal{I}(\nu,u, x)^2\mathcal{P}_\zeta(p)\mathcal{P}_\zeta(|\bm{k}-\bm{p}|)(2\pi)^3\delta(\bm{k}+\bm
{k'}) \ .
\end{align}
Then, using the following relationship
\begin{align}
e^+_{ij}(\hat{\bm k})p_ip_j = \dfrac{p^2}{\sqrt{2}}\sin^2\theta\cos2\phi \ , \qquad e^\times_{ij}(\hat{\bm k})p_ip_j = i\dfrac{p^2}{\sqrt{2}}\sin^2\theta\sin2\phi \ ,
\end{align}
%
%
%
we obtain the power spectrum of induced gravitational waves
\begin{align}
\mathcal{P}_{h, \rm i}(\tau,k) &= \mathcal{P}^{++}_{h, \rm i}(\tau,k) + \mathcal{P}^{\times\times}_{h, \rm i}(\tau,k) \notag \\
&= \dfrac{128}{81x^2}\int_0^\infty d\nu\int_{|1-\nu|}^{1+\nu}du ~\overline{\mathcal{I}^2(\nu, u, x)}\left[\dfrac{4\nu^2 - (1 - u^2 + \nu^2)^2}{4\nu u}\right]^2\mathcal{P}_{\zeta}(k\nu)\mathcal{P}_{\zeta}(ku) \ . \label{eq: ins}
\end{align}
We note that the over-line means a time average of $\mathcal{I}$ since it oscillates much faster than the cosmological time scale after it sufficiently reenters the sub-horizon regime.
For this reason, we should evaluate \eqref{eq: ins} at late times after the horizon crossing.

\subsection{Detectability}

Finally, we calculate the energy spectrum of sourced gravitational waves and discuss their detectability.
The logarithmic energy density of GW at present $\tau = \tau_0$ is given by \cite{Maggiore:1999vm}
\begin{equation}
\Omega_{\rm{GW}}(\tau_0, k) \equiv \dfrac{1}{\rho_c}\dfrac{d\rho_{\rm{GW}}}{d\ln k} \ ,
\end{equation}
where $\rho_c = 3\Mpl^2H_0^2$ is the critical energy density of the present universe.
Since the gravitational waves behave as the radiation after the horizon reentry $\tau = \tau_k$, the energy density of GW today is redshifted as
\begin{equation}
\Omega_{\rm{GW}}(\tau_0,k) = \dfrac{1}{48}\left(\dfrac{a(\tau)k}{a_0^2H_0}\right)^2\mathcal{P}_h(\tau, k) \ , \qquad k = a_kH_k \ ,
\end{equation}
where we consider the time domain $\tau_k \leq \tau \leq \tau_{\text{eq}}$ before the matter-radiation equality and after the horizon-reentry.
Using the entropy conservation law, we get
\begin{align}
	\Omega_\text{GW}(\tau_0,k)
	&\simeq
	\dfrac{0.39}{48}\left( \frac{g_*}{106.75} \right)^{-1/3}
	\left(\dfrac{k}{aH}\right)^2\Omega_{r,0}\mathcal{P}_h(\tau, k) \ ,
\end{align}
where $\Omega_{r,0}$ is the density parameter of radiation at present.

\begin{figure}[htbp]
 \begin{minipage}{0.5\hsize}
  \begin{center}
   \includegraphics[width=75mm]{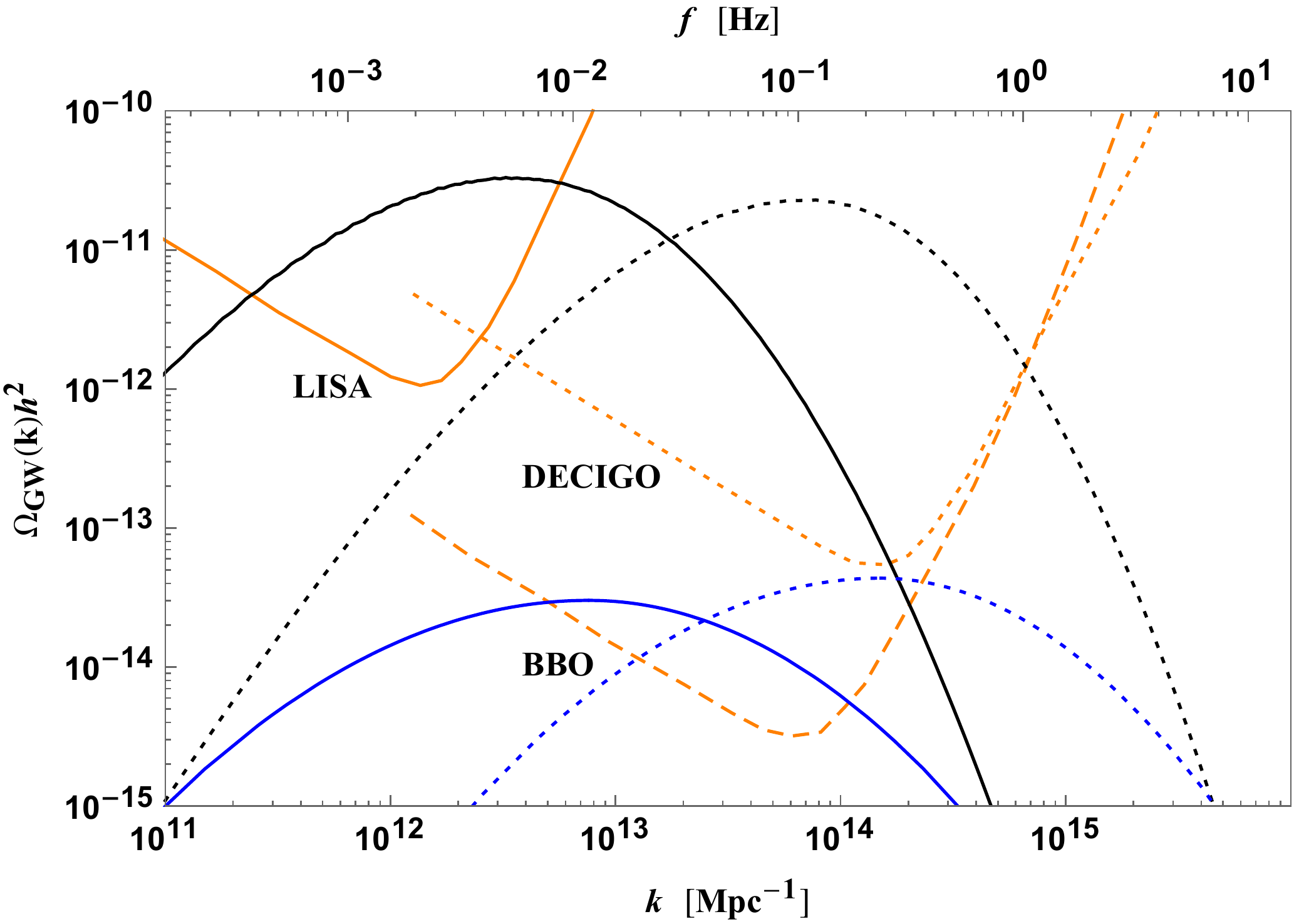}
  \end{center}
 \end{minipage}
 \begin{minipage}{0.5\hsize}
 \begin{center}
  \includegraphics[width=75mm]{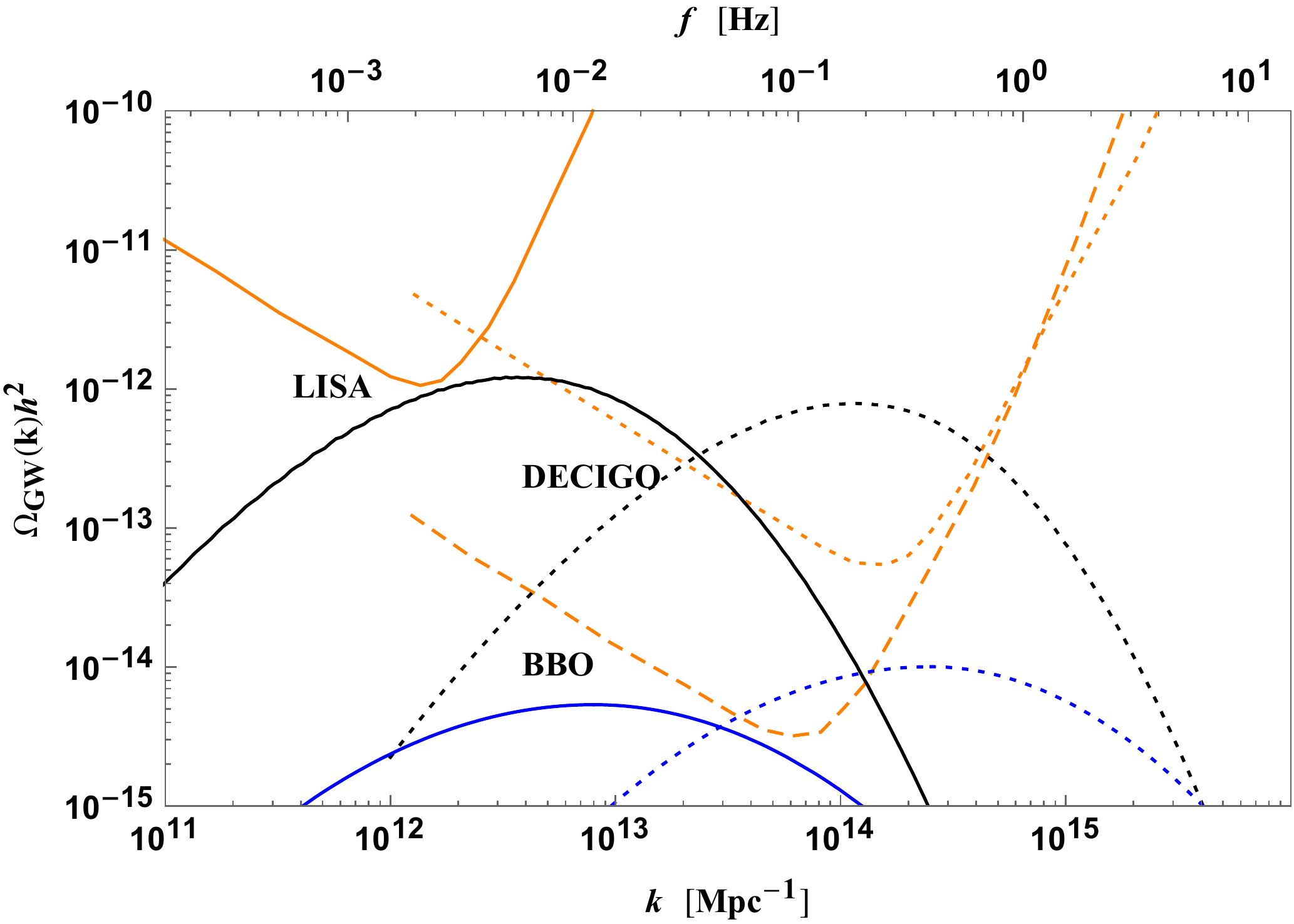}
 \end{center}
  \end{minipage}
\caption{Various populations of gravitational wave power spectrum in our model for the case of Gaussian window function (left panel) and top-hat window function (right panel).
We plot the sensitivity curves of interferometers refered in \cite{Moore:2014lga}.
The black line represents the energy density of induced gravitational waves from Reducible diagram.
The blue line represents the energy density of primordial gravitational waves. 
We plot these spectra with different values of parameter set with solid and dashed line.
}
\label{fig:GWspectru}
\end{figure}

In Figure \ref{fig:GWspectru}, we plot the magnitude of gravitational wave signals, both for the case of the Gaussian window function (left panel) and of the top-hat window function (right panel) in the estimate of PBH mass spectrum as dark matter.
In these graphs, we compare the power spectrum of the induced gravitational waves (black line) and the primordial gravitational waves (blue line) peaked at two different frequencies (solid and dashed lines), corresponding to LISA or DECIGO/BBO interferometer scales.
In any case, the peak amplitude of primordial gravitational waves is always smaller than that of induced gravitational waves.
We find that its suppression factor is roughly given by $\sim(10^{-3}/\mathcal{P}_{\zeta,s})r_{\rm v}^2$ ~evaluated at the time when these spectra become maximized.
We notice that the power spectra of induced gravitational waves here are only originating from the Reducible diagram  so that the total amount of induced gravitational waves will be more detectable. 
For the case of Gaussian window function, the signals of induced gravitational waves are potentially testable with LISA or DECIGO/BBO missions, while those of primordial gravitational waves are challenging to detect by LISA and DECIGO.
On the other hand, for the case of top-hat window function both signals become smaller than those for Gaussian window function and hence are difficult to test with LISA.

We are interested in distinguishing the signature of primordial and induced gravitational waves in terms of their statistical properties such as non-gaussianity.
Although it might not be directly measured \cite{Bartolo:2018evs, Bartolo:2018rku}, we might have a chance to test the squeezed tensor non-gaussianity by means of probing the quadrupolar anisotropy induced in the tensor power spectrum \cite{Dimastrogiovanni:2018uqy, Ozsoy:2019slf, Fujita:2019tov}.
We would like to explore them in future work.

\section{Discussion}
\label{discussion}

In this section, we discuss some possibilities we need to check and ensure our proposal.

\subsection*{A. Black hole binary events}

As the main prediction, we explored the observability of PBH dark matter and the accompanied gravitational waves after inflation.
It will be also worthwhile to search for the possibility of generating PBHs for the gravitational wave binary events \cite{Bird:2016dcv, Clesse:2016vqa, Sasaki:2016jop, Eroshenko:2016hmn, Carr:2016drx}.
In our model setup, however, it would be challenging.
The mass spectrum of PBHs explaining binary events is derived from the curvature perturbation power spectrum with a sharp peak of around $N\sim40$.
In order to construct it, a large $c_1$ value is needed to make $n(t)$ greater than 2 at an earlier stage of inflation.
By doing this, however, the overall amplitude of $n(t)$ simultaneously increases and consequently it leads to an enhancement of curvature perturbation power spectrum with a broad bump.
Such a broad spectrum is prohibited by the observational constraint of CMB $\mu$-distortion \cite{Inomata:2016rbd}.
As a next step, we are going to construct an extended model of this work that predicts binary PBHs with satisfying the above constraint.

\subsection*{B. Backreaction of gauge field}

We evaluate the backreaction of gauge field \eqref{eq: eleenergy} and discuss its effect on the background dynamics in our model.
In \eqref{eq: eleenergy}, the momentum interval is restricted to the window having exited the horizon since the contribution from the sub-horizon regime should be renormalized.
The fractional energy density of any given mode reaches a maximum value around the transition time and in particular the momentum exiting the horizon at around $n(t) = 2$ is mostly contributed in the momentum integration (see Figure \ref{fig:deltaA}).
We denote a momentum scale corresponding to the maximal amplitude as $k = k_m$.
Since $\langle \rho_E(t) \rangle$ cannot be written as the analytical form in our model, we use the following numerical expression of the maximum value of backreaction
\begin{align}
\langle\rho_{E}\rangle_{\text{max}} \simeq \mathcal{A}\left.\dfrac{d\langle\rho_{E}(t_{\rm peak})\rangle}{d\ln k}\right|_{k = k_m} \ ,
\end{align}
where $\mathcal{A}$ is the numerical factor obtained by the momentum integration.
In our parameter set, it gets $\mathcal{A} \sim 5$.
The condition with neglecting backreaction of gauge field to Friedmann equation reads
\begin{equation}
\dfrac{\langle\rho_E\rangle_{\text{max}}}{3\Mpl^2H^2} \ll 1 \ .
\end{equation}
This condition automatically holds if the backreaction is negligible compared to the motion of the inflaton
\begin{equation}
\dfrac{2\bar{I}_\varphi}{\bar{I}}\langle\rho_E\rangle_{\text{max}} \ll 3H\dot{\bar{\varphi}} \qquad \longleftrightarrow \qquad \dfrac{\langle\rho_E\rangle_{\text{max}}}{3\Mpl^2H^2} \ll \dfrac{\epsilon_H}{n_{\rm{max}}}  \label{eq: back} \ .
\end{equation}
We have checked that our parameter sets safely satisfy the above conditions: the value of the left-hand side in \eqref{eq: back} becomes about two orders of magnitude smaller than that of the right-hand side at the relevant time when the particle production occurs.
Even with the backreaction slightly modifies the motion of inflaton, the transition of coupling function occurs at a certain time and therefore our predictions will not be dramatically changed.

\section{Conclusion}
\label{conclusion}

In this work, we studied the phenomenology of particle production from the dilaton-gauge coupling during inflation and the observational signatures on intermediate scales smaller than CMB measurements.
We considered the possibility that the fluctuation of the gauge field is amplified on super-horizon scales due to the background motion of the coupled inflaton field, whose growth power $n(t)$ is characterized by the time variation of the coupling function.
Regarding the functional form of $I(\varphi)$, we adopted the exponential function since it universally appears from the point of view of higher dimensional theory.
In this case, the growing power of gauge field $n(t)$ monotonically increases and the particle production can take place at the late stage of inflation when $n(t)$ is greater than the critical value $n(t) = 2$.
In addition to the exponential type, we also introduced another dilaton-independent term in the coupling function, which may also appear by considering the string-loop expansion effect in powers of the dilaton coupling constant \cite{Damour:1994zq}.
Due to the presence of constant term, $n(t)$ stops increasing at a certain time and makes a transit behavior during inflation and consequently a short-time particle production of gauge field takes place at the intermediate stage of inflation.
We found that the scalar and tensor modes exiting the horizon at around $n(t) =2$ are significantly enhanced by the tachyonic instability of the gauge field.
As a result, the sourced power spectrum becomes a bumpy shape that is peaked at a scale much smaller than Mpc-Gpc.
We used some parameter sets and demonstrated that the enhanced curvature power spectrum predicts the formation of PBHs with mass $10^{17}$-$10^{22} \text{g}$, whose abundance can explain all of the dark matter in our present universe.
Moreover, we also analyzed the primordial gravitational waves provided during inflation and the induced gravitational waves sourced by the second-order curvature perturbation after inflation.
We showed that their amplitudes are potentially testable with the future space-based laser interferometers such as LISA, DECIGO, or BBO missions.

While we have clarified new cosmological signatures from the dilaton-gauge field dynamics during inflation, we also expect that the model of inflation with the two-form field can predict similar observables.
This is because the two-form field can be also amplified due to the slow-roll motion of the scalar field via its kinetic coupling function \cite{Ohashi:2013mka, Ito:2015sxj, Obata:2018ilf}.
We leave these issues in future work.

\section{Acknowledgement}
This work is supported by the JSPS KAKENHI Grants No. 17H01131 (M. K.), No. 17K05434 (M. K.), and No. 19K14702 (I.O.), MEXT KAKENHI Grant No. 15H05889 (M. K.), World Premier International Research Center Initiative (WPI Initiative), MEXT, Japan (M. K., H. N.), Advanced Leading Graduate Course for Photon Science (H.N.), and the JSPS Research Fellowships for Young Scientists Grant No. 19J21974 (H. N.). 

\appendix

\section{Polarization vector and tensor}
\label{ap: p}

Here we discuss the polarization vector and tensor.
The polarization vectors with the wave vector $\hat{\bm{k}} = (\sin\theta\cos\phi, \sin\theta\sin\phi, \cos\theta)$ are defined as
\begin{equation}
e^{X}_i({\hat{\bm{k}}}) = (\cos\theta\cos\phi, \cos\theta\sin\phi, -\sin\theta) \ , \qquad e^{Y}_i({\hat{\bm{k}}}) = (-\sin\phi, \cos\phi, 0)
\end{equation}
in order to obey the transverse and orthogonal relations $k_ie^{\lambda}_i(\hat{\bm{k}}) = 0, \ e^{X}_i(\hat{\bm{k}})e^{X}_i(-\hat{\bm{k}}) = 1, \ e^{Y}_i(\hat{\bm{k}})e^{Y}_i(-\hat{\bm{k}}) = -1, \ e^{X}_i(\hat{\bm{k}})e^{Y}_i(-\hat{\bm{k}}) = 0$.
Moreover, they satisfy the following identities
\begin{align}
&e^X_i(\hat{\bm{p}})e^X_i(\widehat{\bm{k}-\bm{p}}) = -\cos\theta_{\hat{\bm{p}}}\cos\theta_{\widehat{\bm{k}-\bm{p}}} + \sin\theta_{\hat{\bm{p}}}\sin\theta_{\widehat{\bm{k}-\bm{p}}}
= -\cos(\theta_{\hat{\bm{p}}} + \theta_{\widehat{\bm{k}-\bm{p}}}) \label{eq: id1} \ , \\
&e^Y_i(\hat{\bm{p}})e^Y_i(\widehat{\bm{k}-\bm{p}}) = -1 \label{eq: id2} \ , \\
&e^X_i(\hat{\bm{p}})e^Y_i(\widehat{\bm{k}-\bm{p}}) = e^Y_i(\hat{\bm{p}})e^X_i(\widehat{\bm{k}-\bm{p}}) = 0 \label{eq: id3} \ .
\end{align}
In terms of the polarization vectors, we can define the polarization tensors as
\begin{align}
&e^+_{ij}(\hat{\bm k}) = \dfrac{1}{\sqrt 2}\left( e^X_{i}(\hat{\bm k})e^X_{j}(\hat{\bm k}) - e^Y_{i}(\hat{\bm k})e^Y_{j}(\hat{\bm k}) \right) \ , \\
&e^\times_{ij}(\hat{\bm k}) = \dfrac{i}{\sqrt 2}\left( e^X_{i}(\hat{\bm k})e^Y_{j}(\hat{\bm k}) + e^Y_{i}(\hat{\bm k})e^X_{j}(\hat{\bm k}) \right)
\end{align}
which satisfy the transverse-traceless condition.
Assuming $\bm k$ is directed in $\hat{z}$ axis, we get the following identities
\begin{align}
&e^{+}_{ij}(\hat{\bm{k}})e^{X}_{i}(\hat{\bm{p}})e^{X}_{j}(\widehat{\bm{k}-\bm{p}}) = -\dfrac{1}{\sqrt 2}\cos\theta_{\hat{\bm p}}\cos\theta_{\widehat{\bm k - \bm p}} \ , \qquad e^{+}_{ij}(\hat{\bm{k}})e^{Y}_{i}(\hat{\bm{p}})e^{Y}_{j}(\widehat{\bm{k}-\bm{p}}) = \dfrac{1}{\sqrt 2} \label{eq: id4} \ , \\
&e^{+}_{ij}(\hat{\bm{k}})e^{X}_{i}(\hat{\bm{p}})e^{Y}_{j}(\widehat{\bm{k}-\bm{p}}) = 0 \ , \qquad e^{+}_{ij}(\hat{\bm{k}})e^{Y}_{i}(\hat{\bm{p}})e^{X}_{j}(\widehat{\bm{k}-\bm{p}}) = 0 \label{eq: id5} \ , \\
&e^{\times}_{ij}(\hat{\bm{k}})e^{X}_{i}(\hat{\bm{p}})e^{Y}_{j}(\widehat{\bm{k}-\bm{p}}) = -\dfrac{is}{\sqrt 2}\cos\theta_{\hat{\bm p}} \ , \qquad e^{\times}_{ij}(\hat{\bm{k}})e^{Y}_{i}(\hat{\bm{p}})e^{X}_{j}(\widehat{\bm{k}-\bm{p}}) = -\dfrac{is}{\sqrt 2}\cos\theta_{\widehat{\bm k - \bm p}} \label{eq: id6} \ , \\
&e^{\times}_{ij}(\hat{\bm{k}})e^{X}_{i}(\hat{\bm{p}})e^{X}_{j}(\widehat{\bm{k}-\bm{p}}) = 0 \ , \qquad e^{\times}_{ij}(\hat{\bm{k}})e^{Y}_{i}(\hat{\bm{p}})e^{Y}_{j}(\widehat{\bm{k}-\bm{p}}) = 0 \label{eq: id7} \ ,
\end{align}
where we introduced the sign function $s = \pm 1$ in \eqref{eq: id6} which change its sign when we take the Hermitian conjugate. 

\section{Analytical solution of gauge mode function}
\label{ap: a}

Here we analyze the gauge mode function obeying \eqref{eq: gauge2}.
The slow-roll solution \eqref{eq: na} tells us that $n(t)$ before the transit time can be represented by
\begin{equation}
n = \dfrac{b}{\ln x + \tfrac{b}{n(t_k)}} \ , \qquad b \equiv \dfrac{c_1}{M_{\rm Pl}\gamma} \ ,
\end{equation}
where $t_k$ is the time at which $k = a(t_k)H$, corresponding to the conformal time $\tau_k = -1/k$.
Then \eqref{eq: gauge2} is rewritten as
\begin{equation}
\left[\partial_u^2 - \partial_u + e^{2u} - \dfrac{b}{u+\tfrac{b}{n(t_k)}}\left( \dfrac{b}{u+\tfrac{b}{n(t_k)}} - 1 \right)\right](\bar{I}A_k) = 0 \qquad (u \equiv \ln x) \ . \label{eq: u}
\end{equation}
On super-horizon scales (neglecting $e^{2u} \ll 1$), one can find that \eqref{eq: u} can be replaced with the following form
\begin{align}
&\left[N\partial^2_N + ( \beta - N )\partial_N - \alpha \right](N^{-\tfrac{1}{2}\beta}\bar{I}A_k) = 0 \ , \label{eq: chokika} \\ 
&\alpha \equiv \dfrac{1}{2}(1 + \sqrt{1+4b^2}-2b) \ , \qquad \beta \equiv 1+\sqrt{1+4b^2}
\end{align}
in terms of  $N = u + b/n(t_k)$.
This is known as Kummer's equation whose solution is given by the combination of two linearly independent confluent hypergeometric functions of the first kind $M(\alpha, \beta; N)$ and the second kind $U(\alpha, \beta; N)$.
Of these terms, the growing mode on super-horizon scales corresponds to $U(\alpha,\beta; N)$.
Hence we can write the solution as
\begin{align}
&\bar{I}A_k = \dfrac{1}{\sqrt{2k}}N^{\tfrac{1}{2}\beta}C(k)U(\alpha, \beta; N) \ , \label{eq: geo}
\end{align}
where $C(k)$ is an integration constant determined by connecting \eqref{eq: geo} to the numerical solution near horizon-crossing. 

\begin{figure}[tbp]
\center
  \includegraphics[width=90mm]{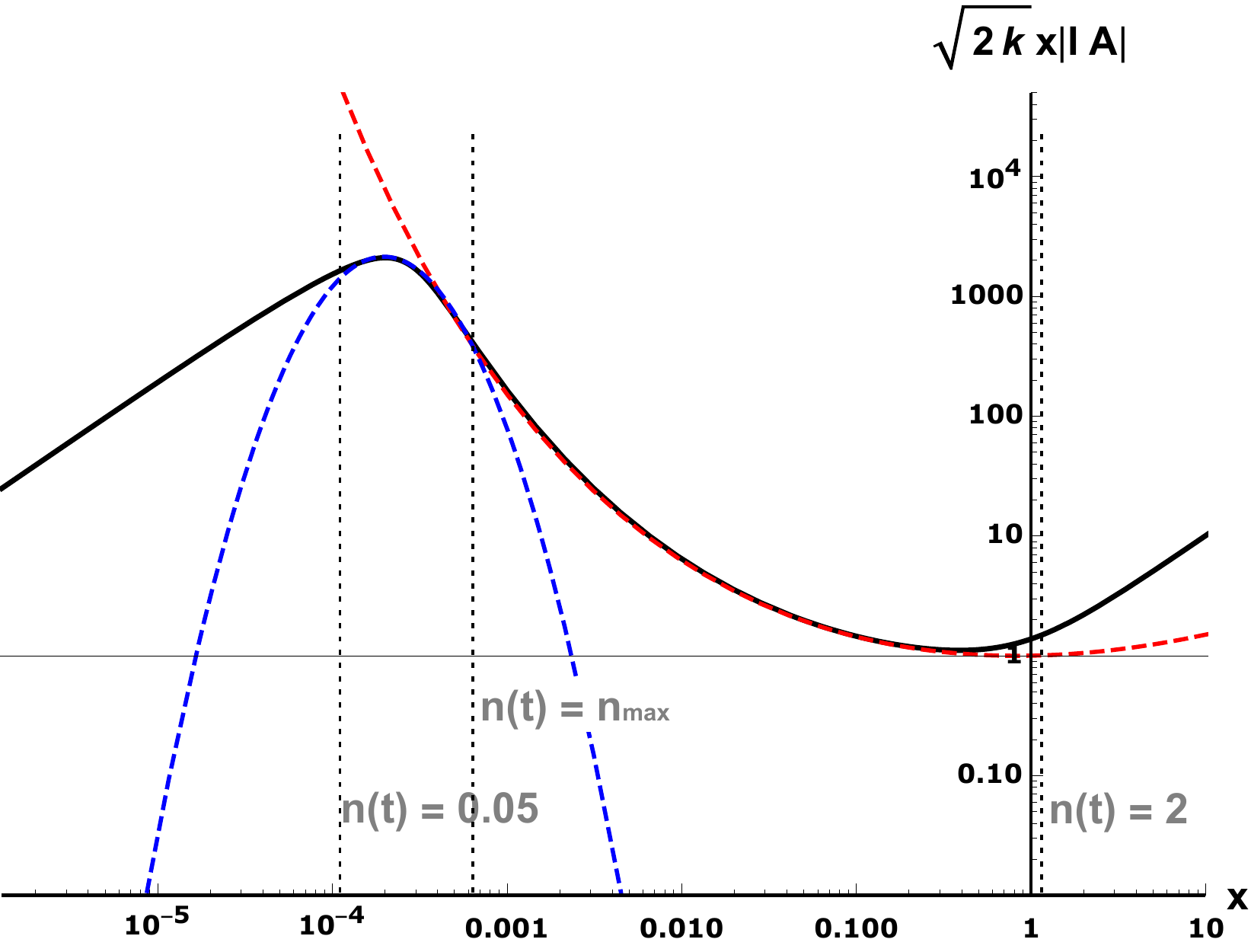}
  \caption
 {
 A time evolution of $\bar{I}A_k$ with the momentum scale crossing the horizon at $N_k \sim 10$.
 The solid black line shows the numerical solution of \eqref{eq: gauge2}.
 The dashed red and blue lines show the slow-roll solution \eqref{eq: geo} and the fitting function \eqref{eq: fit} with $C(k) = (N_k^{\beta/2}U(\alpha, \beta; N_k))^{-1}, \ \sigma^2 = 0.80$.
 We connect them at the time when $n(t) = n_{\text{max}}$.
 These analytical functions well describe the numerical solution until $n(t) \sim 0.05 \ll 2$ .
 }
 \label{fig:deltaAA}
\end{figure}

\begin{figure}[tbp]
\center
  \includegraphics[width=100mm]{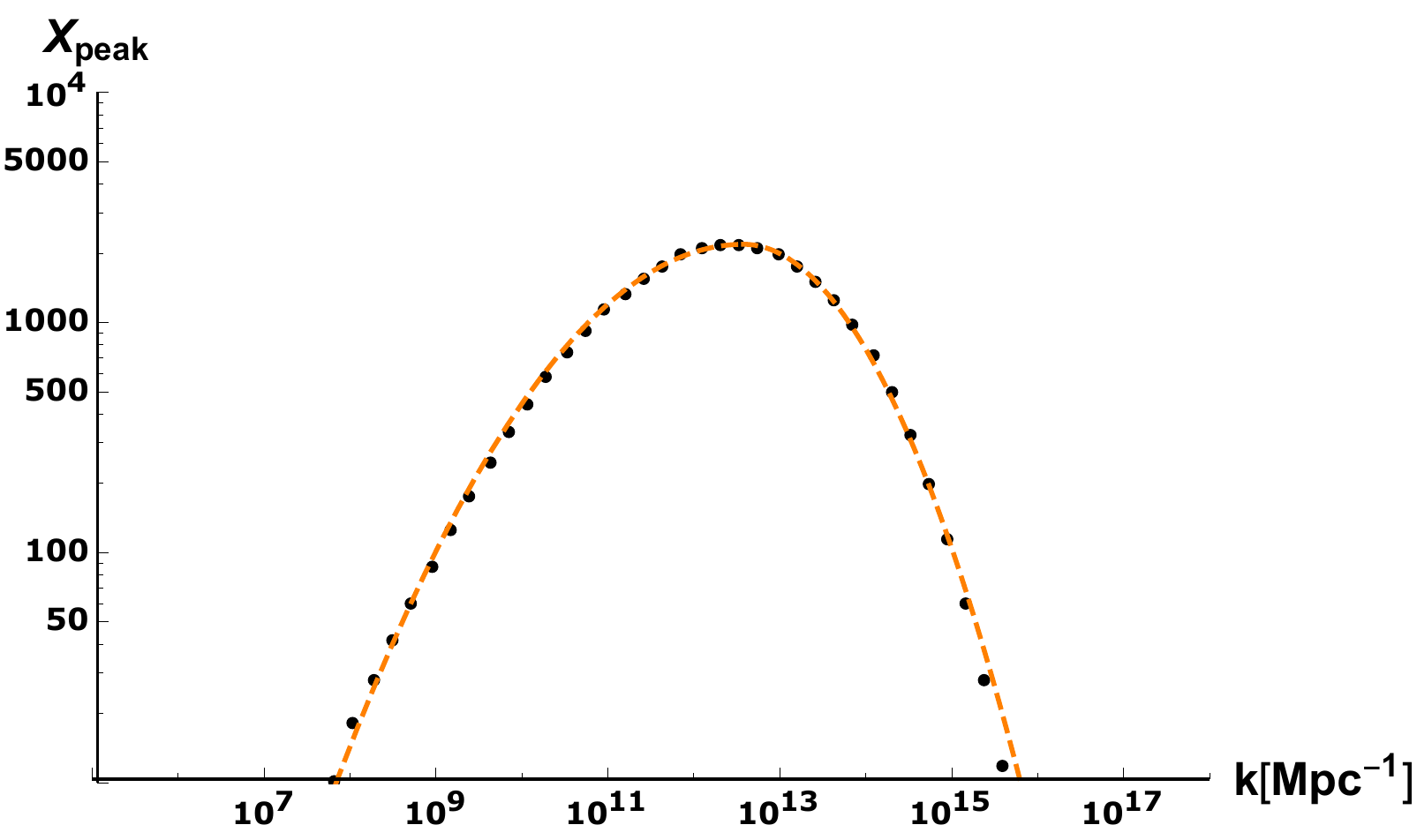}
  \caption
 {
 A plot of $X_{\rm peak}$ with respect to momentum.
 The black dots show the maximum value of numerical solution \eqref{eq: gauge2} with the parameter set $\{B_2/B_1\simeq 8.8\times10^{25}, \ c_1\simeq23 \}$.
 The dashed orange line shows the fitting Gaussian function \eqref{eq: Xk} with $k_p = 3.8\times10^{12}\text{Mpc}^{-1}$ and $\sigma^2_X(k) = 3.2^2\Theta(k - k_p) + 4.7^2\Theta(k_p - k)$.
 }
 \label{fig:Xpeak}
\end{figure}

The analysis of gauge mode function around the transition time is complicated because $n(t)$ is not given by the simple relation of e-folds. 
Instead of seeking the solution written by the closed form, we fit it by using a Gaussian function
\begin{equation}
\bar{I} A_k(x) \simeq \bar{I} A^{\rm fit}_k(x) \equiv \dfrac{1}{\sqrt{2k}x}X_{\rm peak}(k)\exp{\left[-\dfrac{\left(\ln (x/x_{\rm peak})\right)^2}{\sigma^{2}}\right]} \label{eq: fit}
\end{equation}
with the amplitude $X_{\rm peak}(k)$ and variance $\sigma$.
The width of $\sigma$ is simply determined by the background motion of $n(t)$
and therefore does not have an explicit scale-dependence.
We can estimate $X_{\rm peak}(k)$ by connecting \eqref{eq: fit} to \eqref{eq: geo} when $n(t) = n_{\rm max}$:
\begin{align}
X_{\rm peak}(k) &= x_m \exp{\left[\dfrac{\left(\ln (x_m/x_{\rm peak})\right)^2}{\sigma^{2}}\right]}N(x_m)^{\tfrac{1}{2}\beta}C(k)U(\alpha, \beta; N(x_m)) \ . \label{eq: Xk}
\end{align}
In Figure \ref{fig:deltaAA} we depict a time evolution of gauge mode function crossing the horizon at $N_k\sim 10 ~(n(t_k) \sim 2)$ and compare it with the obtained analytical solutions.
We can see that they are well fitted to the numerical solution in the time domain when $n(t)$ is effective for the generation of perturbations.
We also plot $X_{\rm peak}(k)$ in Figure \ref{fig:Xpeak}.
Around the relevant scales, $X_{\rm peak}$ is well described by the following fitting function
\begin{equation}
X_{\rm peak}(k) \simeq A_0\exp\left(-\dfrac{\ln(k/k_p)^2}{\sigma_X^2(k)}\right) \label{eq: fit2} \ .
\end{equation}

%
%

%
%

\end{document}